%% file: scu_date.tex
\documentclass[10pt,journal,compsoc]{IEEEtran}
\IEEEoverridecommandlockouts
\usepackage{cite}
\usepackage{amsmath,amssymb,amsfonts}
\usepackage{algorithm}
\usepackage{algorithmic}
\usepackage{multirow}
\usepackage{booktabs}
\usepackage{graphicx}
\usepackage{textcomp}
\usepackage{hyperref}
\usepackage{xcolor}
\def\BibTeX{{\rm B\kern-.05em{\sc i\kern-.025em b}\kern-.08em
    T\kern-.1667em\lower.7ex\hbox{E}\kern-.125emX}}

\usepackage[american]{babel}

\newlength\figureheight
\newlength\figurewidth

\usepackage{units}

\usepackage[acronym,nomain,notranslate]{glossaries}
        \glsdisablehyper{} 
        \usepackage{relsize}  
        \input{acro.tex}
        
\newcommand{\secref}[1]{Section~\ref{#1}}
\newcommand{\subsecref}[1]{Subsection~\ref{#1}}
\newcommand{\figref}[1]{Fig.~\ref{#1}}
\newcommand{\algoref}[1]{Alg.~\ref{#1}}
\newcommand{\tableref}[1]{Table~\ref{#1}}

\usepackage{pgfplots}
\usepgfplotslibrary{groupplots}
\pgfplotsset{every axis/.append style={
    label style={font=\small},
    title style={font=\small},
    legend style={font=\small},
                    tick label style={font=\small}  
                    }}
                  \usetikzlibrary{patterns}
                  \usetikzlibrary{shapes,arrows,backgrounds,fit,positioning,chains,calc,automata,matrix,trees,decorations}
\pgfplotsset{compat=1.14}                  

\begin{document}

\title{Synapse Compression for Event-Based Convolutional-Neural-Network Accelerators
}

\author{
    \IEEEauthorblockN{Lennart Bamberg\IEEEauthorrefmark{1}, Arash Pourtaherian\IEEEauthorrefmark{2}, Luc Waeijen\IEEEauthorrefmark{2}, Anupam Chahar\IEEEauthorrefmark{1}, and Orlando Moreira\IEEEauthorrefmark{2}}
    \IEEEauthorblockA{\\ \IEEEauthorrefmark{1}At the time of writing with GrAI Matter Labs:
    bamberg@uni-bremen.de, aschahar@gmail.com}
     \\ \IEEEauthorblockA{\IEEEauthorrefmark{2}GrAI Matter Labs:
    \{apourtaherian, lwaeijen, omoreira\}@graimatterlabs.ai}
}


\maketitle

\begin{abstract}
Manufacturing-viable neuromorphic chips require novel compute architectures to achieve the massively parallel and efficient information processing the brain supports so effortlessly.  
The most promising architectures for that are spiking/event-based, which enables massive parallelism at low complexity.
However, the large memory requirements for synaptic connectivity are a showstopper for the execution of modern \glspl{cnn} on massively parallel,  event-based architectures. 
The present work overcomes this roadblock by contributing a lightweight hardware scheme to compress the synaptic memory requirements by several thousand times---enabling the execution of complex \glspl{cnn} on a single chip of small form factor. 
A silicon implementation in a 12-nm technology shows that the technique achieves a total  memory-footprint reduction of up to  \unit[374]{$\times$} compared to the best previously published technique at a negligible area overhead.
\end{abstract}
\glsunset{cnn}


\input{introduction.tex}

\input{related_work.tex}

\input{background.tex}

\input{technique.tex}

\section*{Acknowledgments}
This publication is supported by the \textit{EU Horizon 2020} project,
\textit{ANDANTE}, which has received funding from the \textit{ECSEL Joint Undertaking} (JU) under grant agreement No.~876925. The JU receives support from the \textit{European Union’s Horizon 2020} research and innovation program and France, Belgium, Germany, Netherlands, Portugal, Spain, Switzerland.

\bibliography{scu_date}
\bibliographystyle{IEEEtran}
\section*{Biographies}
\begin{IEEEbiography}
    [{\includegraphics[width=1in]{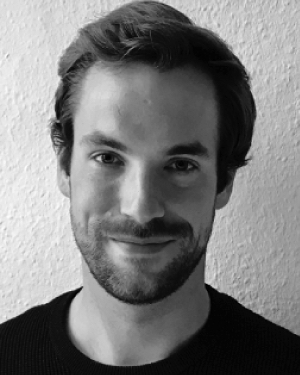}}]{Lennart Bamberg}
 received the B.Sc., M.Sc., and Ph.D. degree with \textit{summa cum laude} in electrical and information engineering from the University of Bremen, Germany, in 2014, 2016, and 2020 respectively. He has been a Research Scholar at Georgia Tech and a Technical Director at GrAI Matter Labs. Currently, he is a Principal Architect at NXP where he is responsible for the company's AI IPs\@. Moreover, he is a lecturer at the University of Bremen.
\end{IEEEbiography}
\begin{IEEEbiography}
    [{\includegraphics[width=1in]{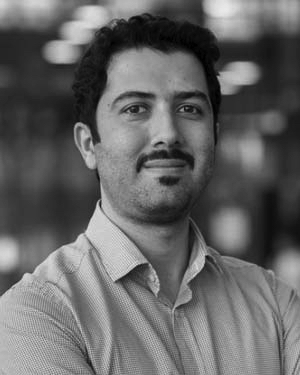}}]{Arash Pourtaherian}
  received the B.Sc. degree in electrical engineering jointly from Indiana University–Purdue University, Indianapolis, IN, USA, and the University of Tehran, Iran. In 2010 and 2018 respectively he received the M.Sc. and Ph.D. degree in electrical engineering from the Eindhoven University of Technology. From 2018 to 2019 he was a post-doc at the Eindhoven University of Technology. Currently, he is a Principal Computer Architect at GrAI Matter Labs.
\end{IEEEbiography}
\begin{IEEEbiography}
    [{\includegraphics[width=1in]{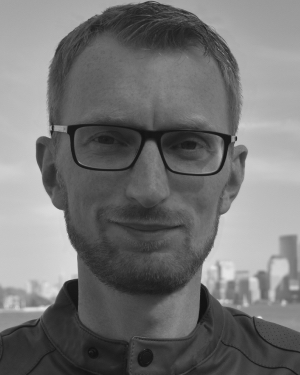}}]{Luc Waeijen}
 received the B.Sc., 
 M.Sc., and Ph.D.~degrees from the Eindhoven University of Technology (TU/e), in 2012, 2013, and 2022, respectively. He is a Senior Computer Architect at GrAI Matter Labs, driving the design and specification of neuromorphic processors. 
\end{IEEEbiography}

\begin{IEEEbiography}
    [{\includegraphics[width=1in]{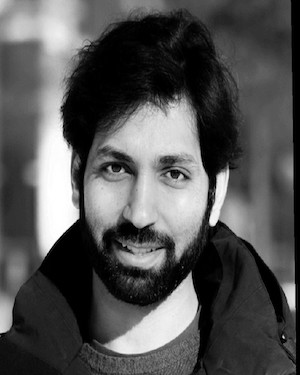}}]{Anupam Chahar}
 received the B.Sc. from Jaypee University of Information Technology in 2010 and the M.Sc. from TU Delft in 2012. Since 2022, he is a
Lead Cryptography Firmware Engineer at PQshield. In the past, he held various software engineering positions at ASML, Intrinsic ID, Intel, and GrAI Matter Labs. 
\end{IEEEbiography}

\begin{IEEEbiography}
    [{\includegraphics[width=1in]{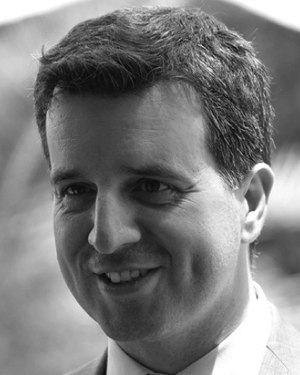}}]{Orlando Moreira}
    received a Ph.D. degree in electrical engineering from the Technical University of Eindhoven, The Netherlands, in 2012. He has been a Research Scientist with Philips Research, a Senior Scientist with NXP Semiconductors, and a Principal Engineer with ST-Ericsson and Intel. He is currently Chief Architect at GrAI Matter Labs, where he is also responsible for the compute architecture.
\end{IEEEbiography}

 \end{document}

%% file: acro.tex
\newacronym{scu}{PEG}{programmable event generator}
\newacronym{dnn}{DNN}{deep-neural-network}
\newacronym{cnn}{CNN}{convolutional neural network}
\newacronym{lut}{LUT}{look-up table}
\newacronym{mac}{MAC}{multiply-accumulate}
\newacronym{fm}{FM}{feature map}
\newacronym{lif}{LIF-NN}{leak-integrate-fire neural network}
\newacronym{ann}{ANN}{artificial neural network}
\newacronym{3d}{3D}{Three-dimensional}
\newacronym{2d}{2D}{two-dimensional}
\newacronym{noc}{NoC}{network-on-chip}
\newacronym{id}{ID}{identifier}
\newacronym{psl}{ESU}{event-to-synapse unit}
\newacronym{soc}{SoC}{sytem-on-chip}
\newacronym{simd}{SIMD}{single-instruction-multiple-data}
\newacronym{ai}{AI}{artificial intelligence}
\newacronym{dram}{DRAM}{dynamic random access memory}
\newacronym{sram}{SRAM}{static random access memory}
\newacronym{finfet}{FinFET}{fin field effect transistor}
\newacronym{tsmc}{TSMC}{Taiwan Semiconductor Manufacturing Company}
\newacronym{dfg}{DFG}{data-flow graph}
\newacronym{gpu}{GPU}{graphical processing unit}
\newacronym{cpu}{CPU}{central processing unit}
\newacronym{sd}{SD-NN}{sigma-delta neural network}
\newacronym{lsb}{LSB}{least-significant bit}
\newacronym{fifo}{FIFO}{first-in first-out}
\glsunset{3d}
\glsunset{2d}
\glsunset{tsmc}
\glsunset{finfet}
\glsunset{id}
\glsunset{sram}
\glsunset{dram}
\glsunset{ai}
\glsunset{gpu}
\glsunset{cpu}
\glsunset{lsb}
\glsunset{soc}
\glsunset{fifo}

%% file: introduction.tex
\section{Introduction}
\Acrfullpl{cnn} achieve state-of-the-art computer-vision performance.
Neural networks are not only more accurate but also easier to develop than hand-crafted techniques. Thereby, they enable a plethora of applications, opening new markets for digital systems.

One of the challenges \glspl{cnn} pose is their  large computational complexity and the massive parameter count (e.g., \textit{ResNet50} requires over \unit[23]{M} parameters and \unitfrac[4]{GFLOPS}{frame}~\cite{he2016deep}). Therefore, \gls{cnn} inference is relatively slow and power-hungry on traditional general-purpose compute architectures, as it requires a large number of expensive \gls{dram} accesses and arithmetic operations. Thus, high frame-rate and yet low-power \gls{cnn} inference is challenging on constrained edge devices. Consequently, accelerators for efficient \gls{cnn} inference  at the edge
are one of the largest research topics in industry and academia today.

Some of the most promising processor architectures for efficient \gls{cnn} inference are dataflow/event-based with a self-contained memory organization~\cite{loihi2:online,moreira2020neuronflow,akopyan2015truenorth}.
The neuromorphic dataflow model of computation, where execution is triggered on the arrival of data exchanged through events rather than by a sequential program, enables scaling to massive core counts at low complexity. 
Moreover, dataflow cores are typically lighter in hardware implementation than more traditional \textit{von-Neumann} processors. 

Another advantage of event-based systems is that they can exploit the large degrees of activation/firing sparsity in modern deep neural networks. This is especially true for leak-integrate-and-fire ({\small LIF}) or sigma-delta neuron models, which exploit temporal correlation in the input and the feature maps to increase sparsity~\cite{o2016sigma,moreira2020neuronflow}. Thus, event-based processing delivers massive compute performance at low power consumption and area. 

A self-contained memory architecture overcomes \gls{dram} access bottlenecks, as all parameters required during execution are stored in local on-chip memories, distributed over the cores. This avoids slow and power-hungry \gls{dram} accesses entirely. The need for no external
\gls{dram} makes the system not only more efficient power and performance wise, but also
cheaper and smaller. This further enhances the suitability of event-based systems for embedded applications.

Due to these vast promises, a range of event-based and self-contained architectures was designed by academia~\cite{moradi2017scalable,furber2014spinnaker}, and global semiconductor companies like \textit{Intel} and \textit{IBM}~\cite{akopyan2015truenorth,davies2018loihi,loihi2:online,moreira2020neuronflow}. However, many of the existing chips only allow a limited neuron count and hence do not support the inference of modern sophisticated \glspl{cnn} (e.g., \textit{ResNet}~\cite{he2016deep} or \textit{MobileNet}~\cite{howard2017mobilenets}).
A fundamental problem is the memory requirement for {synapses}, i.e., the weighted connections between the neurons. 
 Published event-based architectures, store the {synapses} in (hierarchical) \glspl{lut}, resulting in memory requirements that scale with the total neuron count. This leads to high memory requirements for neural networks with a large neuron and synapse count.
 Consequently, architectures like \textit{Intel}'s \textit{Loihi} and \textit{IBM}'s \textit{TrueNorth} are not capable to execute modern, sophisticated \glspl{cnn} as shown in \tableref{tab:neu_syn}.
\begin{table}[]
\centering
\setlength{\tabcolsep}{3.3pt} 
\scriptsize 
\caption{Neuron and synapse count of 
\glspl{cnn} and the maximum capabilities of event-based architectures from major semiconductor companies.}\label{tab:neu_syn}
\begin{tabular}{lccc|cc}
\toprule
                 & \multicolumn{3}{c|}{CNNs}                                                                 & \multicolumn{2}{c}{Architecture}                           \\
                 & \multicolumn{1}{c}{\textit{PilotNet}~\cite{bojarski2016end}}                 & \multicolumn{1}{c}{\textit{MobileNet}~\cite{howard2017mobilenets}} & \multicolumn{1}{c|}{\textit{ResNet50}~\cite{he2016deep}} & \multicolumn{1}{c}{\textit{IBM}~\cite{akopyan2015truenorth}}            & \multicolumn{1}{c}{\textit{Intel}~\cite{davies2018loihi,loihi2:online}} \\
                 
\cmidrule{1-1} \cmidrule{2-4} \cmidrule{5-6}
\textbf{Neurons}  & {\unit[0.2]{M}} & \unit[4.4]{M}                          & \unit[9.4]{M}                          & \unit[1.1]{M} & \unit[1.1]{M}                            \\
\textbf{Synapses} & {\unit[27]{M}}   & \unit[0.5]{B}                           & \unit[3.8]{B}                          & \unit[0.3]{B}  & \unit[0.1]{B} \\     
\bottomrule
\end{tabular}
\end{table}
 Sophisticated \gls{cnn} models are still out of reach for existing event-based architectures, limiting their usability today.
 
This work overcomes this severe limitation. It contributes a technique that systematically exploits the {translation invariance} of \gls{cnn} operations to make the synaptic-memory requirements independent of the neuron count. This results in extreme compression. 
Our technique employs two lightweight hardware blocks, a \gls{scu} and an \gls{psl}\@. The \gls{scu} computes the synaptic connectivity of thousands of neurons---grouped in a \textit{neuron population}---using a single instruction word, the \textit{axon}. The \gls{psl} fills the synaptic connections with weights. Weights can be shared among all neurons of a population, reducing the memory requirements for parameters.


\begin{figure}
    \centering
    \includegraphics[scale=0.22]{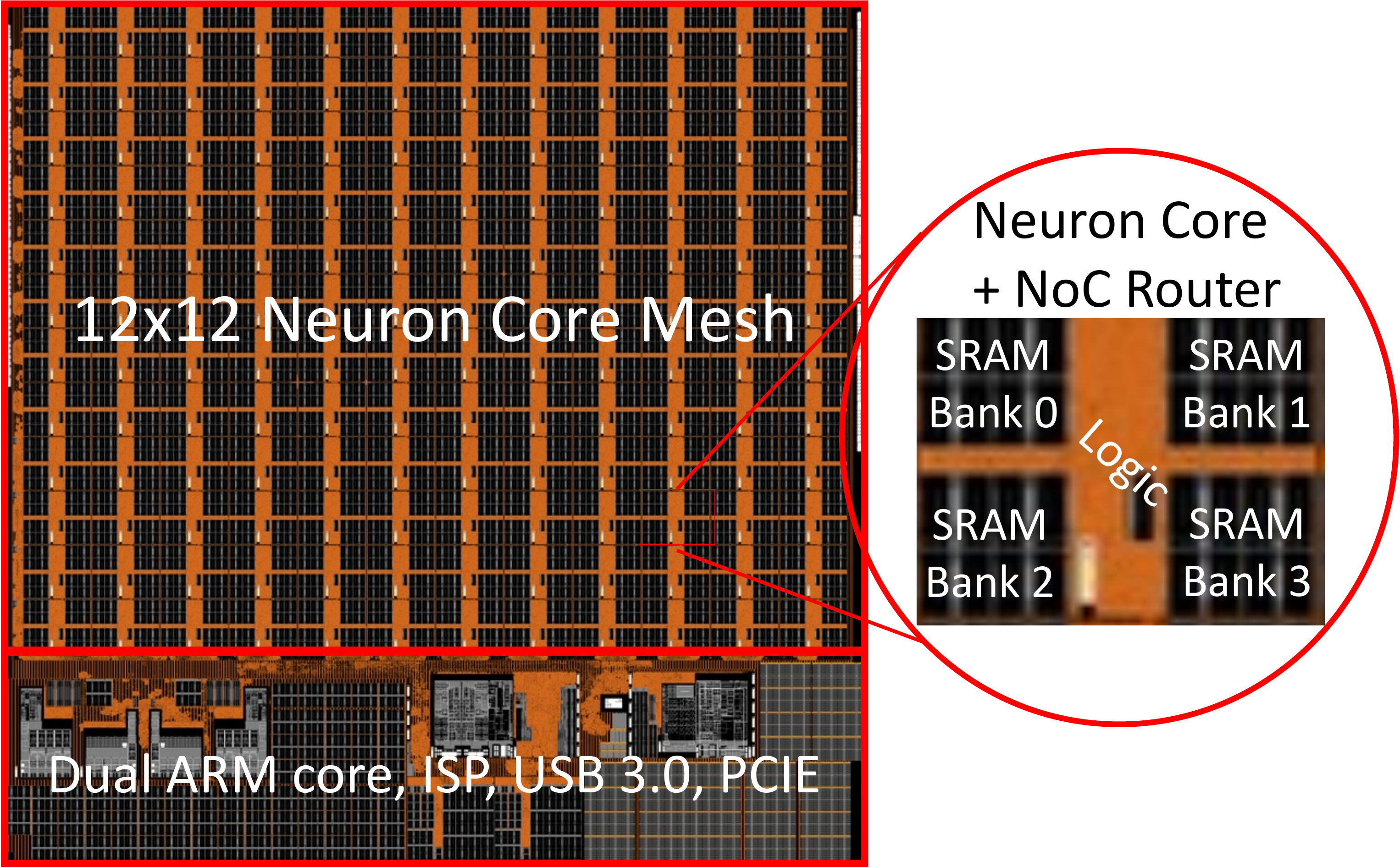}
    \caption{Taped-out 144-neuron-core \acrshort{soc} with \acrshort{simd} execution.}
    \label{fig:die_shot}
\end{figure}

Our proposed technique was taped-out 
through the  144-neuron-core \textit{GrAI-VIP} shown in \figref{fig:die_shot}.
In the 12-nm \gls{soc} of size \unit[60]{mm$^2$}, our technique exhibits negligible implementation cost, while providing overall memory compression rates of over \unit[300]{$\times$} for modern \glspl{cnn} compared to the best-published reference technique.

\textit{Intel} recently revealed its new self-contained and event-based \gls{dnn} accelerator \textit{Loihi 2}. It is  capable to run \textit{Nvidia}'s \textit{PilotNet} \acrshort{cnn} in \unit[30]{mm$^2$}
using its 7-nm technology~\cite{loihi2:online}.
The same network (i.e., without any extra optimizations) requires only 3 out of 144 cores on our chip using a larger technology node at 2{$\times$} the die area.
This demonstrates the practical value of the proposed technique, as it allows event-based cores to run much more sophisticated/complex \glspl{cnn} in a temporal-sparse fashion on a single low-cost, self-contained, and event-based accelerator.

The rest of this paper is structured as follows. Related work is discussed in \secref{sec:related_work}.
\secref{sec:prelim} includes the preliminaries covering \glspl{cnn} and their sparse execution on event-based accelerators. The proposed technique is derived in \secref{sec:tech}. \secref{sec:exp_res} includes discussions and experimental results. Finally, a conclusion is drawn.

%% file: related_work.tex
\section{Related Work}\label{sec:related_work}
This section reviews related work on event-based architectures and memory-footprint compression.
A wide range of massively parallel, self-contained, event-based architectures
have been proposed (e.g.,\cite{moradi2017scalable,furber2014spinnaker,davies2018loihi,moreira2020neuronflow,akopyan2015truenorth,loihi2:online,BraiChip:online}). Most only support neuromorphic neuron models, while \textit{GrAI Matter Lab}'s \textit{Neuronflow} architecture~\cite{moreira2020neuronflow} and the \textit{Loihi2}~\cite{loihi2:online} support also traditional \glspl{dnn}---executed in a time-sparse manner to reduce the compute requirements~\cite{o2016sigma,moreira2020neuronflow}. Such \glspl{dnn} are today more accurate and easier to~train.

Reducing the memory footprint of neural networks is an actively researched topic. Compression is vital for self-contained architectures, as non-fitting \glspl{cnn} cannot be supported.
Traditional architectures have looser memory constraints due to the usage of external \gls{dram}\@. However, even here compression has great advantages, as it reduces latency- and power-hungry \gls{dram} accesses~\cite{han2015deep}.

One approach to reducing memory requirements at the application level is to use a lighter-weight \gls{cnn} (e.g.,~\cite{xie2017resnext,howard2017mobilenets}).
Another possibility is to execute a given \gls{cnn} at a lower precision. A common technique is to run the network in 8-bit integer arithmetic~\cite{jacob2018quantization}, but even binary network implementations exist~\cite{rastegari2016xnor}. 
\textit{Han et al.} proposed to iteratively prune weights with a small magnitude (effectively setting them to 0), followed by incremental training to recover accuracy~\cite{han2015deep}. In combination with 8-bit integer arithmetic and weight entropy-encoding, weight pruning can achieve compression rates larger than \unit[10]{$\times$} for modern \glspl{cnn}~\cite{han2015deep}.
 
The approaches mentioned above do not only reduce the memory footprint, but also the computational complexity of the \gls{cnn} inference. Also, since these techniques transform the neural network itself, they are effective for most compute architectures. The drawback is that the techniques are \textit{lossy}; i.e., they typically entail some accuracy loss that is partially recoverable  by retraining~\cite{han2015deep,jacob2018quantization}.

Synapse-compression techniques for event-based systems, such as the one proposed in this work, 
 are \textit{orthogonal} to the \textit{lossy} techniques discussed above. Hence,
both compression types can be applied in conjunction and studied independently. 
Since synapse-compression techniques are \textit{lossless}, they do not affect the network prediction/output. Hence, these techniques, tailored for event-based systems, entail no loss in network accuracy. Thus, they can be applied post-training.

 
 Early event-based architectures used a single routing \gls{lut} in the source to look up all synaptic connections of a neuron on its firing~\cite{moradi2017scalable}.
 This approach results in many events and large memory requirements for \glspl{dnn} with many neurons, $|\mathcal{N}|$, 
 as 
 the number of required bits is of complexity $\mathcal O(|\mathcal N|\log_2(|\mathcal N|))$. 
\textit{Moradi et al.}\ and \textit{Davies et al.}, independently, found that the events can be reduced and the synaptic-memory compressed to $\mathcal O(|\mathcal N|\sqrt{\log_2(\mathcal |\mathcal N|)})$
 through a hierarchical, tag-based routing scheme~\cite{moradi2017scalable,davies2018loihi}. This hierarchical routing scheme found application in \textit{Intel's} original \textit{Loihi} architecture and the \textit{DYNAPs} architecture. 
 
 The technical note of the recently released \textit{Loihi 2} claims that the synaptic memory footprint for \glspl{cnn} has been compressed again by up to \unit[17]{$\times$}~\cite{loihi2:online}. 
 Unfortunately, details of the approach are (at the time of writing) not fully disclosed. However, at a reduction of up to 17$\times$, the synaptic memory requirements must still be at least of complexity $\mathcal O(|\mathcal N|)$. Thus, larger \glspl{cnn} would remain unfeasible for \textit{Loihi2}. 


The idea of generating a synaptic memory structure for event-based \gls{cnn} inference that scales independently of $|\mathcal N|$ (required to run today's complex \glspl{cnn}), has been tried for the \textit{SpiNNakker} chip~\cite{serrano2015convnets}. However, the existing method is not generic enough. It only supports simple convolution layers.
Other common \gls{cnn} operations such as stride-2 convolutions, polling, and depth-wise convolutions are all not supported. Moreover, the technique does not allow the cutting of \glspl{fm} in multiple fragments for compute or memory load-balancing. 
Hence, the technique was only demonstrated in Ref.~\cite{serrano2015convnets} to work for a tiny \gls{cnn} with five standard convolution layers and a maximum feature shape of $28 \times 28$. Thus, it cannot be implemented for modern deep \glspl{cnn} which include many more layer types. Moreover, modern \glspl{cnn} have large \gls{fm} dimensions and strongly heterogeneous compute requirements per layer which make \gls{fm}-cuts inevitable. Thus, the problem addressed by this work remained unsolved until this point.


%% file: background.tex
\section{Preliminaries}\label{sec:prelim}
\subsection{Neural Networks}\label{sec:nn}

 In neural networks for image processing, a \textit{neuron population} is represented by a three-dimensional matrix, $\mathbf{P} \in \mathbb{R}^{D \times W \times H}$. The last two dimensions determine the \gls{2d} Cartesian-grid position to correlate a neuron with a location in space; the first coordinate indicates the feature (or channel) that is sampled. Thus, a single neuron population represents multiple features. Here, the number of channels of such a multi-channel \gls{fm} is denoted as $D$. The \gls{2d} submap of channel $c_i$ ($\mathbf{P}_{c,x,y}|_{c=c_i}$) has the dimensions of a single feature, defined by a tuple: width, $W$, and height, $H$.
 
The raw input image represents the input population of the network.
 For example, a $300\times 300$ \textit{RGB} image is represented by a $3 \times 300 \times 300$ $\mathbf{P}$ matrix, where $\mathbf{P}_{c,x,y}$ is the pixel value for color channel $c$ at location ($x$, $y$).
Starting with this input population, new \glspl{fm} are extracted out of the existing \glspl{fm} successively (feed-forward structure). One such extraction step is symbolized as a network layer in a dataflow graph structure. The most intuitive form is a fully connected layer, in which each neuron in the destination \gls{fm}, $\mathbf{P^+}$, is a function of all neurons in a source \gls{fm},~$\mathbf{P}$: 
\begin{equation}\label{eq:fully_connected}
\mathbf P^+_{c,x,y} = \sigma \Bigg( \sum_{i=0}^{D{-}1}\sum_{j=0}^{W{-}1}\sum_{k=0}^{H{-}1} \mathbf W_{c,x,y,i,j,k}\mathbf P_{i,j,k}+b_c \Bigg) 
\text{.}
\end{equation}
Here, $b_c$ is the bias of the neurons of feature $c$, and $\mathbf{W}$ is the weight matrix. A \gls{dnn} can be implemented in its traditional form or as a spiking \gls{lif}\@.
In the first variant, $\sigma()$ is a non-linear activation function to enable non-linear feature extraction. Often, a rectifier/\textit{ReLU} is used, clipping all negative values to zero. The $\mathit{ReLU}$ function exhibits low hardware complexity. Neuromorphic \glspl{lif} differ from traditional \glspl{dnn} through a more complex, stateful activation, $\sigma ()$, with periodic state-leakage.

A major problem with fully connected layers is the large number of synaptic weights which determine the compute and memory requirements. For example, extracting ten $300\times 300$ features from a $3\times 300\times 300$ \gls{fm} requires \unit[243]{M} weights and \gls{mac} operations. 

This can be overcome by exploiting \textit{spatial locality}~\cite{zhang2021dive}. For a meaningful feature extraction, it is not needed to look far in XY to glean relevant information for one XY coordinate of the extracted feature. Thus, outside a relative range of XY size ($\mathit{KW}$,\,$\mathit{KH}$)---denoted as the receptive field---all synapses between neurons are removed. This drastically reduces the compute and memory requirements for such \textit{locally connected} \gls{2d} layers.

Moreover, network filters should respond similarly to the same patch, irrespective of its XY location in the source \gls{fm}\@. This principle is called \textit{translation invariance}~\cite{zhang2021dive}. Hence, the relative weights can be independent of the XY location of the destination neuron; i.e., weights can be reused for all neurons of the same output channel. This further reduces the memory requirements.

The regular convolution operation in a \gls{cnn} adds both, spatial locality and translation invariance compared to fully connected layers. This results in the following formula for a convolution layer with a kernel size of $\mathit{KW}\times \mathit{KH}$:
\begin{equation}\label{eq:conv_no_padding}
\mathbf P^+_{c,x,y} = \sigma \Bigg( \sum_{i=0}^{D-1}\sum_{j=0}^{\mathit{KW}{-}1}\sum_{k=0}^{\mathit{KH}{-}1} \mathbf W_{c,i,j,k}\mathbf P_{i,x+j,y+k}+b_c \Bigg) 
\text{.}
\end{equation}

\begin{figure}
    \centering
    \includegraphics[scale=0.39]{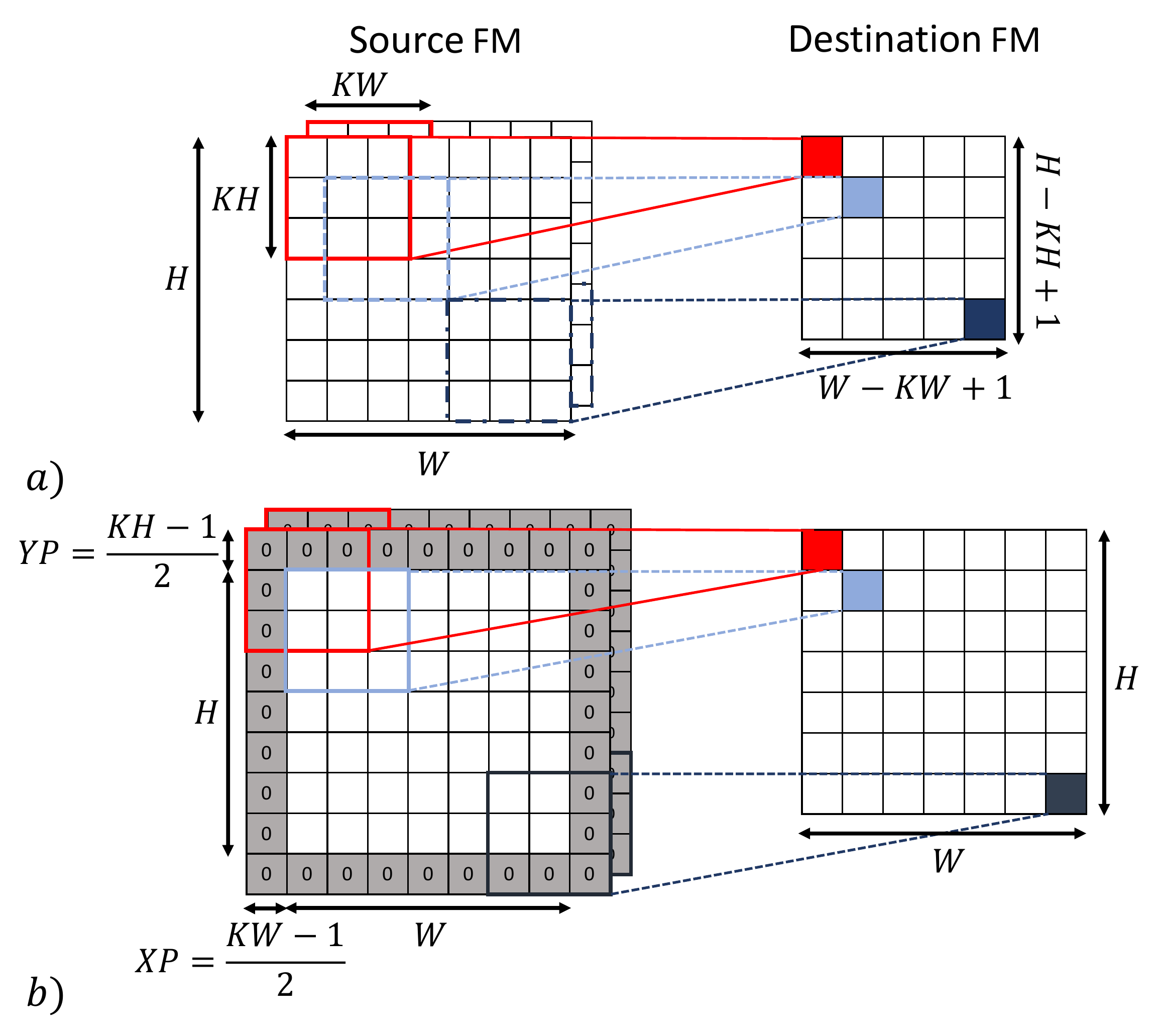}
    \caption{Convolution operation without padding (a), and with "same" zero-padding (b) resulting in no resolution reduction.}
    \label{fig:padding_illus}
\end{figure}

The XY size of the extracted \gls{fm} in a convolution operation is the XY size of the source \gls{fm} minus $\mathit{KW}{-}1$ in width and $\mathit{KH}{-}1$ in height. This is the number of times the convolution kernel fits in the source \gls{fm}, illustrated in \figref{fig:padding_illus}.a.
Padding rows and columns of zeros around the source is done to control the size of the output \gls{fm}\@. To achieve an output of the same size as the input (i.e., to preserve the feature resolution), one must pad $\nicefrac{(\mathit{KH}-1)}{2}$ rows of zeros at the top and bottom 
of the source \gls{fm}, and $\nicefrac{(\mathit{KW}-1)}{2}$ columns at the left and right. \figref{fig:padding_illus}.b illustrates this zero-padding.

Instead of explicitly padding the source \gls{fm}  in a pre-processing step, the effect of zero padding can be expressed implicitly, by changing the source-\gls{fm} indices in Eq.~\eqref{eq:conv_no_padding}:
\begin{align}\label{eq:conv_padding}
P^+_{c,x,y} &= \sigma \Bigg( \sum_{i=0}^{D-1}\sum_{j=0}^{\mathit{KW}{-}1}\sum_{k=0}^{\mathit{KH}{-}1} \mathbf W_{c,i,j,k} \mathbf P_{i,x+j-\mathit{XP},y+k-\mathit{YP}} \nonumber \\ &+b_c  \Bigg)
 \text{\quad with $\mathbf P_{i,x,y}=0$ for $x<0$ or $y<0$}
\text{.}
\end{align}
Here, $\mathit{XP}$ and $\mathit{YP}$ are the rows/columns of zeros padded at the left and the top of the source \gls{fm}, respectively.

Kernel striding and source upsampling are other ways of controlling the dimensions of the extracted features. 
With a stride of two, the kernel is not shifted by one step for every generated output point but by two, resulting in extracted features with a \unit[50]{\%} smaller resolution.
When the source is upsampled to increase the resolution of the extracted features, zeros are filled between the upsampled source values followed by a trainable convolution on the upsampled map. 

\subsection{Event-Based Architectures}
In this section, we outline the conceptual idea of  event-based compute architectures for efficient \gls{cnn} inference. 
These architectures exhibit massive parallelism with many identical cores connected through a \gls{noc}~\cite{moradi2017scalable,loihi2:online}. 

To each of the $C$ cores, one or more neuron populations, $\mathbf{P}_i$, are statically mapped.
Near-memory computing is applied in the architectures. 
Thus, the entire memory is distributed equally over the cores, and the neural parameters (e.g., weighted synapses) are stored in the cores to which the respective neurons are mapped. 
The dynamic variables, computed by one core (i.e, the activated neuron states; see Eq.~\eqref{eq:conv_padding}), and needed as inputs by other cores for the calculation of neuron states, are exchanged through events over the \gls{noc}\@. Hence, there are no coherence problems. 
This allows the removal of higher/shared memory levels, improving latency and power consumption.

A drawback is that local on-chip memories have limited capacity. Thus,
 the static/compile-time mapping of neuron populations to cores must satisfy hard memory constraints; i.e., the number of neurons and synapses that can fit in a core is limited by the core's memory capacity. If a neuron population $\mathbf{P}_i$ of the \gls{cnn} has too many parameters or neurons, it needs to be split/cut into multiple smaller populations before core assignment.

The arrival of data in form of an event triggers the compute  (dataflow execution). 
A buffer stores the received events for \gls{fifo} processing. 
 Event processing multiplies the event value with the weights of the synapses between the source neuron and the destination neurons mapped to the core. The results are accumulated in the destination neuron states.
After accumulation, activated neuron states are sent through events to all synaptically connected neurons in the following layers. Since these neurons can be mapped to any core in the cluster, events can also be injected into the local event queue instead of going out on the \gls{noc}\@.

 
 Note that above we describe how more recent architectures (considered in this paper)
 work. In some earlier architectures, the multiplications of activations by synaptic weights are executed at the source (i.e., before event transmission). Here, event processing at the destination only requires the addition of the event value to the target neuron state ~\cite{moradi2017scalable}. This scheme results in  more events for neurons with multiple synapses. Thus, it is no longer applied in modern architectures. 

\subsubsection{Sparsity Exploitation in Event-Based Architectures} 
 During \gls{dnn} execution, no events are generated for zero-valued activations. Skipping zero-valued events induces no neuron-state accumulation error. 
 Thus, each zero in an \gls{fm} reduces the compute and communication requirements---saving energy and compute resources~\cite{aimar2018nullhop}. This is an advantage of event-based architectures compared to more traditional architectures. Particularly for modern \glspl{cnn} with large degrees of activation sparsity (i.e., zero-valued activations), it gives event-based architectures a power-performance advantage.
 Standard \glspl{cnn} using \textit{ReLU} activations exhibit an activation sparsity of over \unit[50]{\%}, as a \textit{ReLU} clips all negative values to zero~\cite{zhu2022arts}. 
Thus, event-based processing improves the power-delay product by about~\unit[4]{$\times$} at no accuracy loss.

Event-based architectures exhibit even larger gains by implementing \glspl{cnn} as \glspl{lif} or \glspl{sd} to increase sparsity. 
  A {\small LIF}-based \gls{cnn} has the same synaptic connectivity as a standard \gls{cnn} implementation~\cite{cao2015spiking}. 
However, \glspl{lif} exhibit a more complicated firing behavior---expressed by $\sigma()$ in Eq.\ \eqref{eq:conv_padding}. This increases sparsity but also makes back-propagation training hard. 
Consequently,  \glspl{lif} cannot yet compete accuracy-wise with standard \glspl{cnn} for complex applications. 

To address this, \gls{sd} implementations increase the neural sparsity by transforming temporal correlation of neuron potentials between subsequent inferences into event sparsity~\cite{o2016sigma}. This is particularly efficient for time-continuous input data such as video streams.  
By keeping memory for a persistent accumulator state for each neuron, the temporal deltas in the activations can be exchanged/processed instead of absolute values. Still, the same inference results as a standard \gls{cnn} implementation are obtained as state accumulation is a linear operation. Hence, \glspl{sd} exhibit no training or accuracy issues. Any standard \gls{cnn} can be executed as an \gls{sd} at no accuracy loss and without retraining. Still, \glspl{sd} improve sparsity noticeably. The temporal deltas of the activations exhibit more zeros than the absolute activation maps, due to the present temporal correlation---especially if activations are quantized~\cite{o2016sigma}. Thereby, \glspl{sd} increase the gains of event-based processing without sacrificing any network accuracy.

The proposed, as well as the reference techniques, work for all three described \gls{cnn} variants (standard, {\small LIF}, and {\small SD}). Thus, in the remainder of this paper, we  uses the term \textit{firing neuron} to describe a neuron of any type for which the synaptic connectivity has to be triggered due to a non-zero activation, activation-delta, or spike. 

\subsubsection{Synaptic-Memory in Existing Architectures}
Event-based accelerators need information on the synaptic connectivity of a firing neuron to identify which neurons to update with which weights.
Mathematically, the connectivity of neurons can be described as a directed graph $\mathcal G =(\mathcal{N},\mathcal S)$, where $\mathcal N$ and $\mathcal S$ are the sets of neurons and synapses, respectively. Each synapse, $s_i \in \mathcal S$, can be described by a tuple $(n_\text{src},n_\text{dst},w)$, where $(n_\text{src},n_\text{dst})\in \mathcal N^2$ is the synapse's source and destination neuron, respectively, and $\mathit{w}$ is the synaptic weight. 

In early event-based accelerators, \glspl{lut} store the synapse information, mapped to the cores together with the neurons.
On the firing of a neuron, the addresses of the destination neurons together with the weights/parameters are looked-up, and for each pair an event carrying the firing value multiplied by the weight is emitted towards the destination neuron.
If $F$ is the average number of outgoing synapses per neuron,
the required memory bits to store the connectivity (i.e., synapses without the weights) for $|\mathcal N|$ neurons in this scheme is 
\begin{equation}\label{eq:neurons_flat}
\mathit{Mem}_\text{c} = |\mathcal{N}|F\log_2{(|\mathcal N|)}\text{.}
\end{equation}
With $B$-bit weights, the parameter requirements become
\begin{equation}\label{eq:weights_flat}
\mathit{Mem}_\text{p} = |\mathcal{N}|F B\text{.}
\end{equation}
For modern neural networks with many neurons,
a \gls{lut} results in high memory requirements. As an illustrative example, let us investigate the last 3$\times 3$ convolution of the \textit{ResNet50} \gls{cnn}~\cite{he2016deep}, with $512 \times 7\times 7$ neurons in the source and destination. Stroing the connectivity of all neurons in this layer in a \gls{lut} requires at least \unit[110]{MB}. The weights  would need an additional \unit[201]{MB} at an 8-bit precision.

Moreover, the standard \gls{lut}-based approach has the disadvantage that it
results in one event per synapse---so $F$ events per firing neuron. 
A hierarchical/two-stage \gls{lut} layout reduces the number of events and the memory requirements~\cite{moradi2017scalable,davies2018loihi}. 
For each spiking neuron, only the addresses of the destination cores are looked up in the first routing table at the source. Thus, only one event is generated per destination core instead of one per destination neuron (while hundreds of neurons are mapped to a single core).
An event contains a $K$-bit tag, stored also in the \gls{lut} at the source, and the firing value. In the destination core, the weights and the target neurons are selected based on the tag
through a second \gls{lut}\@.

For a maximum of $F_{\max}$ in-going synapses per neuron, 
the tag-width must be at least $\log_2(F_{\max})$. 
With $C$ cores, $M$ neurons per core, and an average of $F$ in-going synapses per neuron, the number of bits to store the connectivity of $|\mathcal N|$ neurons reduces to
\begin{align}
    \mathit{Mem}_\text{c} &= \mathit{Mem}_\text{c,src} + \mathit{Mem}_\text{c,dst} \\
    & = 
    |\mathcal N|\Big(\nicefrac{F}{M}\log_2{(F_{\max}C)}+\log_2{(F_{\max})F}\Big)
    \text{,} \nonumber
\end{align}
in the best case.
However, the memory requirement still scales with $|\mathcal N|$. Thus, the scheme is still impractical for modern \glspl{cnn}\@. 
Analyzing again the last $3\times 3$ convolution layer of \textit{ResNet50}, the synaptic memory requirement is still at least \unit[167]{MB}.


%% file: technique.tex
\section{Technique}\label{sec:tech}
\begin{figure}
    \centering
    \includegraphics[scale=0.35]{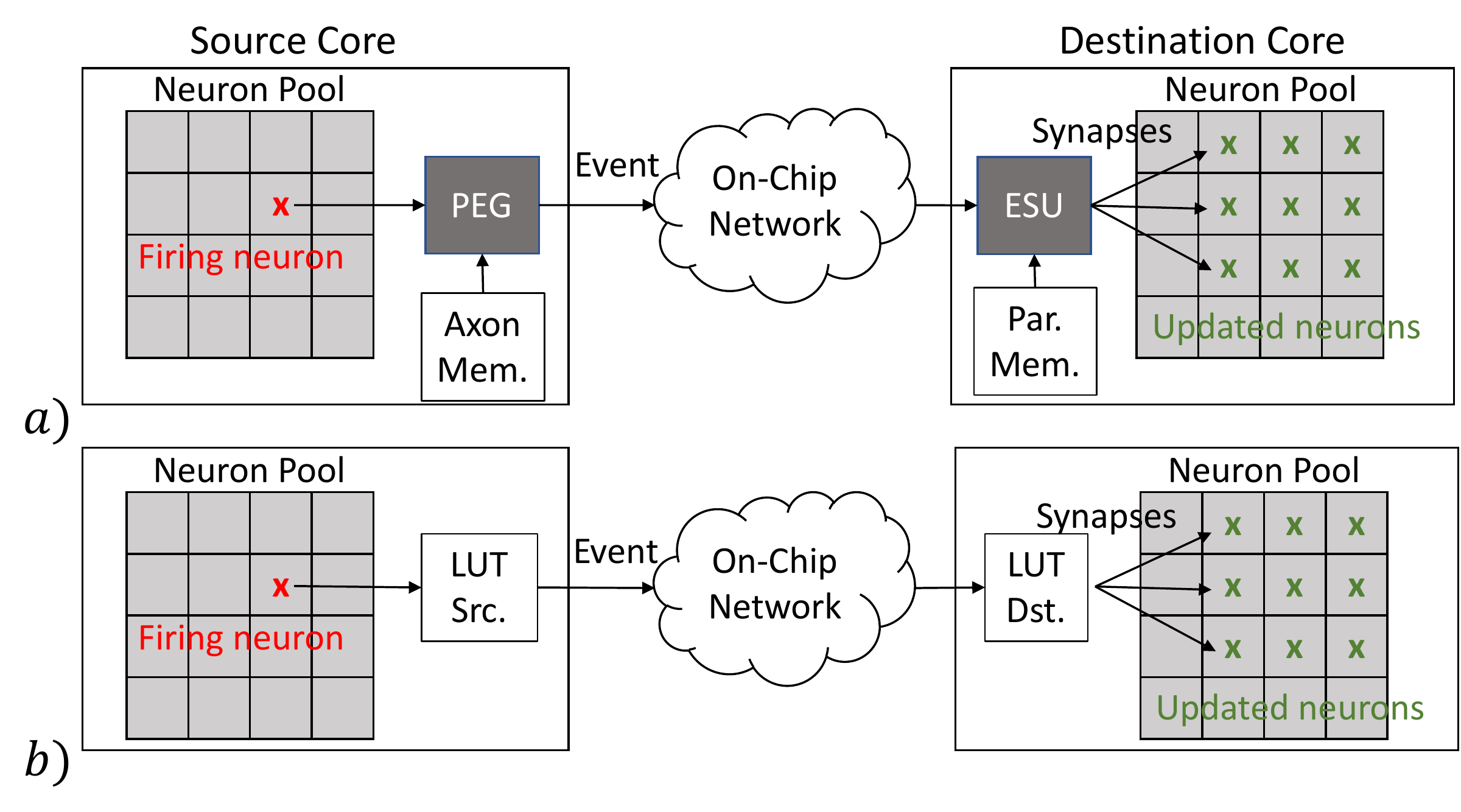}
    \caption{$a$) Proposed technique based on two hardware blocks. $b)$ Reference technique based on a hierarchical \gls{lut}.}
    \label{fig:fund_idea}
\end{figure}

In the following, we propose a technique to compress the synapses for \glspl{cnn}
such that the memory requirements no longer scale with the neuron count but just with the population count.
\figref{fig:fund_idea} illustrates the idea compared to the existing hierarchical-\gls{lut} scheme.
We propose to use two hardware blocks, a \acrfull{scu} and an \acrfull{psl} rather than \glspl{lut}\@. 

To avoid memory requirements per neuron, so-called \textit{axons} in the source connect entire neuron populations 
instead of individual neurons. 
For each population, $\mathbf{P}_{\text{dst},i}$, with at least one synapse directing from the current population, $\mathbf{P}_{\text{src},i}$,
an axon is stored at the source.
 {Axons} are the \gls{scu}'s "instructions", defining simple routines executed on the firing of any neuron to generate the required events. Besides the \textit{axon(s)} of the population, the \gls{scu} needs information on the firing neuron.
Like the reference technique~\cite{moradi2017scalable}, our technique results in low \gls{noc} bandwidth requirements, as the \gls{scu} generates at most one event per {axon}. 

The \gls{psl} in the destination core decodes the event into weighted synapses. The \gls{psl} uses weight sharing between different synapses. This means that events that originate from the same \gls{fm}, but from different neurons, can use the same weights without having to store them multiple times. This yields a weight compression on top of the savings provided by the {axon} scheme due to the translation invariance of \gls{cnn} operations.

In the remainder of this section, we successively derive the architecture of the  \gls{psl} and \gls{scu} to support state-of-the-art \glspl{cnn} at a maximized compression rate. 
We start by looking at the simple case of a regular convolution. First, we assume that the compute and memory capacity of each core is large enough that it is not needed to split any multi-channel \acrshort{fm} into several populations mapped onto different cores.
Afterwards, the approach is extended to cover \gls{fm} fragmentation, kernel striding, and source upsampling.  

\subsection{Basic Technique}
\begin{figure}
    \centering
    \includegraphics[scale=0.36]{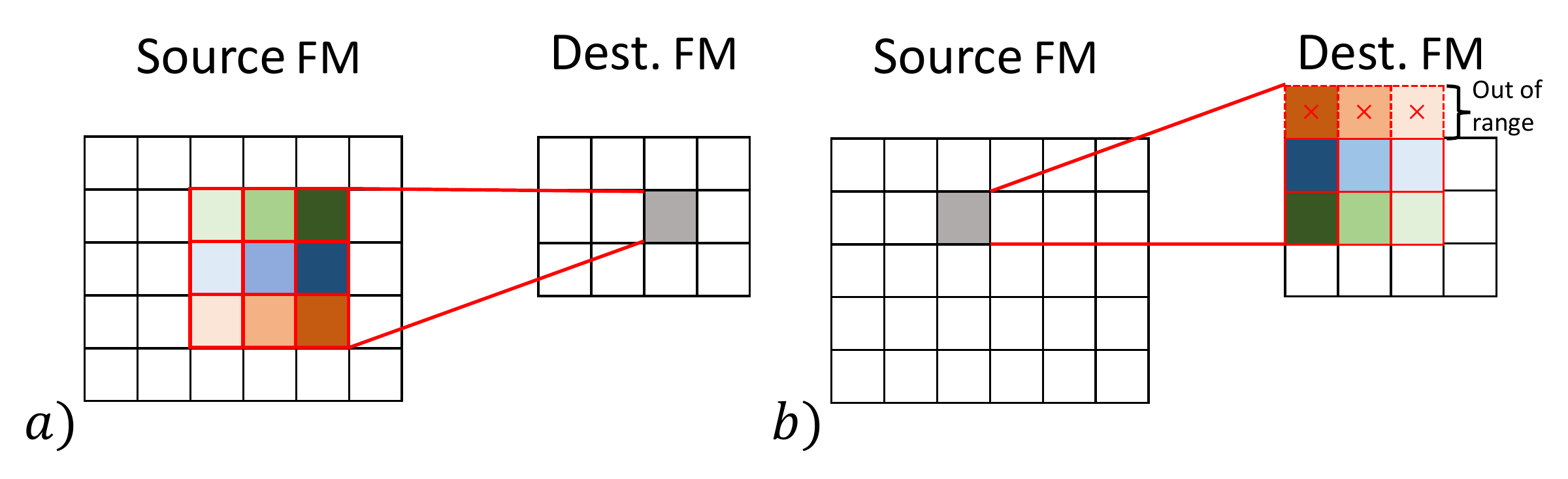}
    \caption{Regular ($a$) and event-based ($b$) convolution implementation.}
    \label{fig:eb_conv}
\end{figure}

 To identify which outgoing synapses a firing neuron has, we must view the convolution operation in a transposed, i.e. event-based, fashion.
 The traditional view (illustrated in \figref{fig:eb_conv}.a for a single channel) reduces source values via the kernel; the event-based view broadcasts source elements via the transposed kernel (illustrated in \figref{fig:eb_conv}.b). 
 Event processing adds the event value multiplied by the respective weight to any neuron state in the range of the transposed kernel. 
 The final neuron states for the inference are obtained after all events have been processed. 

To obtain the same final neuron states in the event-based implementation as in the standard implementation, the weight arrangement for the broadcasted kernel must be transposed in XY (e.g., top-left weight becomes bottom-right) compared to the regular view. 
In the regular view, the destination neuron's XY coordinate is the upper-left anchor point of the kernel in the source-\acrshort{fm}. Thus, in the event-based variant, the XY location of the broadcasted source neuron is the bottom-right anchor point of the transposed kernel in the destination \gls{fm}\@. 
 Equation \eqref{eq:conv_padding} on page \pageref{eq:conv_padding} shows that padding shifts the XY coordinate of the bottom-right anchor-point of the transposed kernel by the padding amount, $(\mathit{XP},\mathit{YP})$.

Thus, on the firing of a neuron on channel $c_\text{src}$ with XY coordinate ($x_\text{src},y_\text{src}$),
 all neurons in the destination \gls{3d} \gls{fm} must be updated whose X and Y coordinate is in $(x_\text{src}$$-$$\mathit{KW}$$+$$\mathit{XP},\, x_\text{src}$$+$$\mathit{XP}]$
and $(y_\text{src}$$-$$\mathit{KH}$$+$$\mathit{YP},\, y_\text{src}$$+\mathit{YP}]$, respectively. In that range, the XY-transposed kernel weights $\mathbf W_{i,j,k,l}|_{l=c_\text{src}}$ are scaled (i.e., multiplied) by the firing value and added to the respective neuron states.
For some firing neurons nearby the edges of an \gls{fm}, part of the broadcasted transposed kernel does not overlap with the destination \gls{fm}, as illustrated in \figref{fig:eb_conv}.b. Such kernel parts located outside the \gls{fm} boundaries do not result in neuron updates. Thus, they are skipped in the \gls{psl}\@.

\begin{algorithm}[t]
\caption{ $\acrshort{scu}$  for  a regular convolution.}
\label{alg:scu_simp}
\begin{algorithmic}[1]
\footnotesize
\REQUIRE {Firing neuron \& value: $x_\text{src}$, $y_\text{src}$, $c_\text{src}$, $v$}
\REQUIRE {Axon set of population: $\mathcal A$}
\FOR{$A = (X_\text{off}, Y_\text{off}, \mathit{AD}_\text{c}, \mathit{ID}_\text{p}) \in  \mathcal A$ }
\STATE $x_{\min},y_{\min} = (x_\text{src},y_\text{src}) + (X_\text{off},Y_\text{off})$
\STATE \textit{gen\_event}($\mathit{AD}_\text{c}, \mathit{ID}_\text{p}, x_{\min}, y_{\min}, c_\text{src}$)
\ENDFOR
\end{algorithmic}
\end{algorithm}
\begin{algorithm}[t]
\caption{\acrshort{psl} for a regular convolution.}
\label{alg:psl_simp}
\begin{algorithmic}[1]
\footnotesize
\REQUIRE Event body:  $x_{\min}, y_{\min}, c_\text{src}, v$
\REQUIRE Neuron population \& shape:  $\mathbf{P}, D, W, H$
\REQUIRE Transposed weights \& kernel size:  $\mathbf{W}$, $\mathit{KW}$, $\mathit{KH}$
\FOR{$\Delta x \in [0,\, \mathit{KW})$}
\STATE $x= x_{\min} + \Delta x$
\STATE \textbf{continue if} $ x \not\in [0,\, W)$ {\color{red}~~// kernel part out of range?}
\FOR{$\Delta y \in [0,\, \mathit{KH})$}
\STATE $y= y_{\min} + \Delta y$
\STATE \textbf{continue if} $y \not\in [0,\, H)$ {\color{red}~~// kernel part out of range?}
\FOR{$c \in [0,\, D)$}
\STATE { \color{red} // update\_neuron depends on neuron model}
\STATE \textit{update\_neuron}($\mathbf P_{c,x,y}$, $\mathbf W_{c,\Delta x, \Delta y, c_\text{src}}$,  $v$)
\ENDFOR
\ENDFOR
\ENDFOR
\end{algorithmic}
\end{algorithm}
The pseudo codes in \algoref{alg:scu_simp} and \algoref{alg:psl_simp} describe the required functionality of the \gls{scu} and the \gls{psl}, respectively. Hardware implementation is done most efficiently through simple state machines. 
The \gls{scu} of the source core (i.e., the core with the firing neuron) calculates the top-left anchor point for the transposed convolution kernel in the destination \gls{fm}, ($x_{\min}$, $y_{\min}$). As outlined above, this requires a subtraction of ($\mathit{KW}$$-$$\mathit{XP}$$-$$1$, \, $\mathit{KH}$$-$$\mathit{YP}$$-$$1$) from the XY coordinates of the firing neuron ($x_\text{src}$, $y_\text{src}$). Since the  padding and kernel shape are constant at \acrshort{cnn}-inference-time, this is done in the \gls{scu} at once by adding a signed offset pair to the neuron's coordinates:
\begin{equation}
    X_\text{off},\, Y_\text{off} = (-\mathit{KW}+\mathit{XP}+1, \, -\mathit{KH}+\mathit{YP}+1)\text{.}
\end{equation}
Storing compile-time calculated offset values rather than the kernel shape and the padding in an {axon}, saves additions/subtractions in the \gls{scu}
for each firing neuron as well as bits in the {axon}.

The outgoing event generated by the \gls{scu} contains the computed ($x_{\min}$, $y_{\min}$) coordinates, which can contain negative values if the top-left anchor-point is located outside the \gls{fm} range. Moreover, the firing value, $v$, and the channel of the firing neuron, $c_\text{src}$, are added to the event body.
Finally, events need the \gls{noc} address of the core to which the target population is mapped, $\mathit{AD}_\text{c}$, as well as a destination-population identifier, $\mathit{ID}_\text{p}$. The later enables  that multiple populations can be mapped onto the same core. These last two parameters are static and thus part of  the {axons}. 

Each source population has an independent {axon} for each directly connected destination population. On each firing, the \gls{scu} must process all these {axons} to generate events towards all populations.

At the far end, the \gls{psl} takes the event and selects the required compile-time-transposed \gls{3d} kernel, $\mathbf W_{i,j,k,l}|_{l=c_\text{src}}$ based on $c_\text{src}$ and $\mathit{ID}_\text{p}$ from the event. It applies the weights scaled by the event value on the neurons with the top-left anchor point taken from the event ($x_{\min}$, $y_{\min}$). It skips kernel parts that are outside the population boundaries. The iteration order is channel-first (\textit{HWC}). This minimizes the amount of required \textit{continue} statements to at most $\mathit{KH}\cdot \mathit{KW}$, reducing the complexity of a hardware implementation.
 
The proposed technique shares kernels and {axons} among all neurons of a population. Hence, besides {axons}, neurons, and weights, we need two more types of memory words: population descriptors and kernel descriptors. A population descriptor contains the \gls{3d} neuron population shape, the neuron type incl.\ the activation function, the start address of the state values, and the number of {axons} linked to the population. A kernel descriptor contains the kernel shape, its weight width, and a memory pointer for the weights.

\subsection{Support for FM Cutting}\label{sec:technique_frag}
 A multi-channel \gls{fm} in the original/pre-mapped \gls{cnn} graph can have so many parameters or neurons that it cannot be mapped onto a single core due to memory or compute constraints. 
Thus, the proposed technique supports the cutting of \glspl{fm} into several  smaller neuron populations, mapped onto different cores. 

An \gls{fm} cut that divides the channels results in a reduction in the number of neurons as well as weights mapped to each core. On the other hand, a cut in the XY space of the features requires all populations to contain all weights. This is due to the translation invariance of the convolution operation. Hence, a cut reducing only the X or Y size of the individual populations increases the overall memory requirements,
as shared weights are duplicated (i.e., mapped onto multiple cores). Nevertheless, our approach supports both: cuts in channel coordinate $C$ and XY. The rationale is that, for high-resolution features, it may happen that even a single channel does not fit into a core, making XY cuts inevitable. 
Also, addressing limitations can result in inevitable XY cuts.  For example, for 8-bit population-width/height fields, a high-resolution \gls{fm} has to be split into populations with a width and height of at most 255.\footnote{Such fragments can be mapped to the same core to share weights.}

\begin{figure}
    \centering
    \includegraphics[scale=0.4]{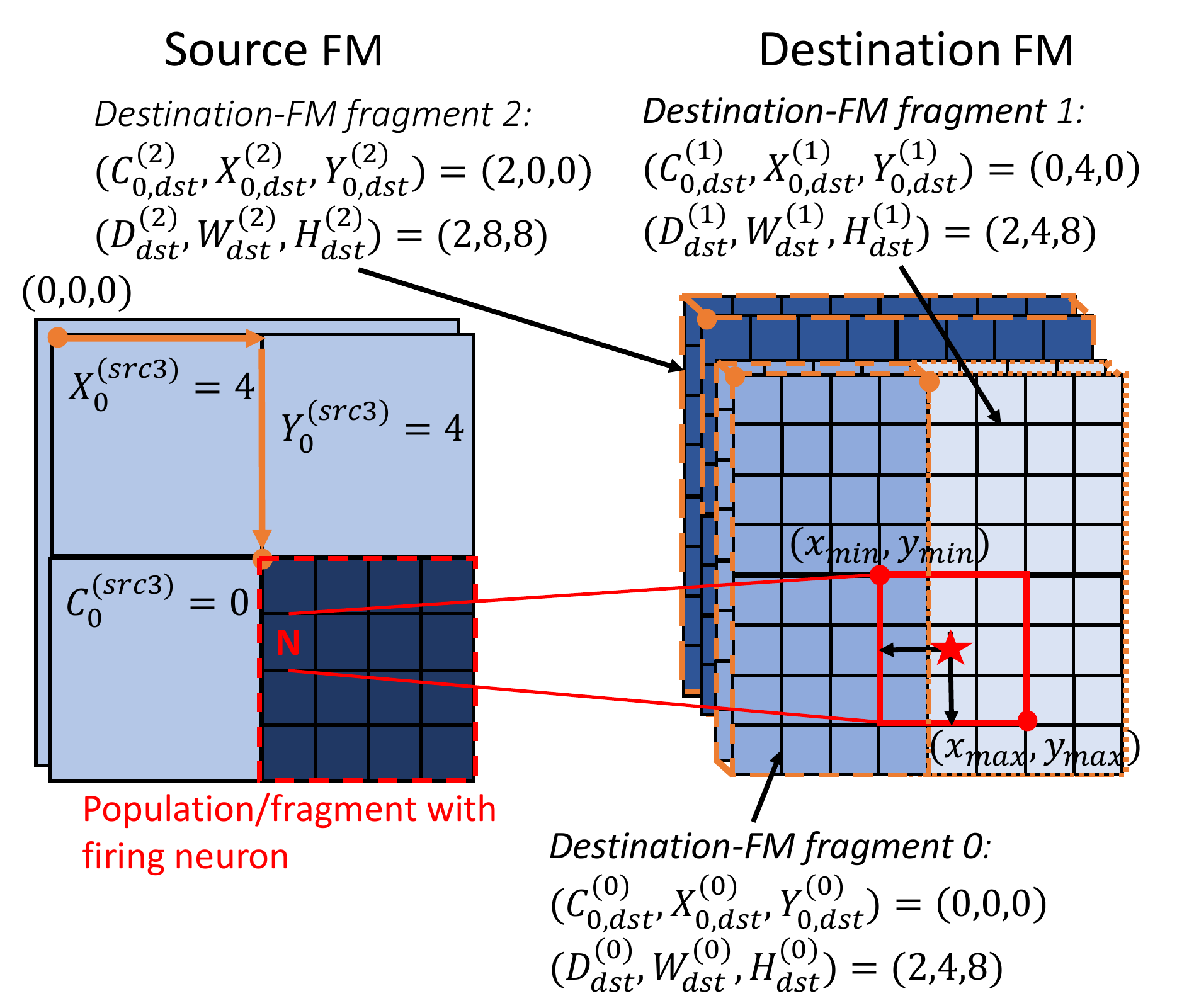}
    \vspace*{-0.2cm}
    \caption{Cutting of the source and the destination \gls{fm}\@.}
    \label{fig:fm_cut}
\vspace*{-0.2cm}
\end{figure}

\figref{fig:fm_cut} shows an example where the source and destination \gls{fm} are cut into four and three pieces, respectively.
The example helps to understand the following mathematical derivation of the \gls{fm}-fragmentation technique. 

Let the shape of the original, unfragmented $i^\text{th}$ \gls{fm} be ($D_{i}, W_{i}, H_{i}$). The \gls{fm} is cut into $M$ disjoint populations $\mathbf{P}_i^{(0)}, \mathbf{P}_i^{(1)}, \dots ,  \mathbf{P}_i^{(M-1)}$. The shape of the $j^\text{th}$ fragment is ($D_i^{(j)},W_i^{(j)},H_i^{(j)}$) and it contains all neurons from the original \gls{fm} with 
$c \in [C_{0,i}^{(j)},\,C_{0,i}^{(j)}$$+$$D_i^{(j)})$,
$x \in [X_{0,i}^{(j)},\,X_{0,i}^{(j)}$$+$$W_i^{(j)})$,
and 
$y \in [Y_{0,i}^{(j)},\,Y_{0,i}^{(j)}$$+$$H_i^{(j)})$.
Thus, $(C_{0,i}, X_{0,i}, Y_{0,i})$ is the coordinate of the upper-left first neuron of the fragment.
For a valid fragmentation, each neuron must be mapped to exactly one population. This is satisfied if the neuron-population fragments are disjoint and if the number of neurons in the pre-mapped \gls{fm} is the same as the sum of the number of neurons of all generated fragments.

Next, we discuss how to adapt the \gls{scu} to cope with \gls{fm} cuts. When a neuron fires in a population, its coordinates, $(c_\text{src}, x_\text{src}, y_\text{src})$, are the ones relative to the upper-left cut point in the original \gls{fm}, not the entire \gls{fm}\@. Thus, adding to the neuron coordinates this start point of the fragment in the original \gls{fm}, $(C_{0,\text{src}},X_{0,\text{src}},Y_{0,\text{src}})$, results in the neuron coordinates in the original/uncut \gls{fm}\@. Hence, the upper-left anchor point of the broadcasted kernel in the original destination \gls{fm} and the channel of the firing neuron are
\begin{align}
    c_\text{src,orig} &= c_\text{src} + C_\text{$0$,src} \nonumber \\
    x_\text{min,orig} &= x_\text{src} + X_\text{$0$,src}- \mathit{KW} + \mathit{XP} + 1 \nonumber \\
    y_\text{min,orig} &= y_\text{src} + Y_\text{$0$,src}- \mathit{KH} +  \mathit{YP} + 1 \text{.}
\end{align}

For each  fragment of the destination \gls{fm}, we then calculate the location of this upper-left anchor point of the transposed kernel relative to their XY start-point in the original \gls{fm}:
\begin{align}
    x_\text{min} &= x_\text{min,orig} - X_\text{$0$,dst} \nonumber \\  &= x_\text{src} + X_\text{$0$,src}- \mathit{KW}  + \mathit{XP} + 1 - X_\text{$0$,dst}  \nonumber \\
    y_\text{min} &= y_\text{min,orig} - Y_\text{$0$,dst} \nonumber \\  &= y_\text{src} + Y_\text{$0$,src}- \mathit{KH} + \mathit{YP} + 1 - Y_\text{$0$,dst} 
    \label{eq:y_min_cutted}
\text{.}
\end{align}
These $(x_\text{min},y_\text{min})$ coordinates,
together with $c_\text{src,orig}$, are sent via an event to the core of the destination-\gls{fm} fragment. 

All parameters but the coordinates of the firing neuron in Eq.~\eqref{eq:y_min_cutted} are constants at run-time. Thus, supporting \gls{fm} cuts just needs an adjustment of the $X$ and $Y$ offset values calculated at compile time, plus an additional offset value for the channel coordinate:
\begin{align}
    C_\text{off} &= C_\text{$0$,src} \nonumber \\
    X_\text{off}  &= X_\text{$0$,src}- \mathit{KW} +  \mathit{XP} + 1 - X_\text{$0$,dst} \nonumber \\
    Y_\text{off} &= Y_\text{$0$,src}- \mathit{KH} +  \mathit{YP} + 1 - Y_\text{$0$,dst} \text{.}
\end{align}
Note that the offset values for X and Y can be negative here. Thus, they are signed values. The channel offset is always positive. Hence, it can be an unsigned value.

For each population/\gls{fm}-fragment pair that is connected by at least one synapse, an {axon} is needed.
However, the number of {axons} remains minimal, as different populations are mapped onto different cores. Thus, different {axons} are required anyway due to the different core addresses.

Each core only stores the weights for the channels that are mapped to it. Thus, the indices of the weight matrix are relative to the first channel of the respective fragment; i.e., when neurons of the channels 4 to 7 are mapped onto a core, $\mathbf W_{0,0,0,0}$ stored in the local core memory is the upper-left kernel weight for source channel 0 and destination channel 4 of the original transposed weight matrix. This fragmentation of the weight matrix is done entirely at compile time.\footnote{All weight sets are only disjoint for no X or Y cuts.}

In summary, besides an additional {axon} parameter $C_\text{off}$  added to the channel coordinate, nothing is added to the \gls{scu} hardware. The \gls{psl} must not be modified at all.

There is an optional extension to the \gls{scu}\@. A \textit{hit detection} to filter out events that an \gls{psl} decodes in zero synapses as the kernel does not overlap at all with the destination population.
Consider in \figref{fig:fm_cut}, that the neuron at position (0,1,1) instead of (0,0,1) fires.  The \gls{scu} depicted by \algoref{alg:scu_simp} sends an event to the core, destination-\gls{fm} fragment 0 is mapped onto. This event has an $x_\text{min}$ value of 4. However, the population only contains neurons with $x \in [0,\,3]$. Thus, event decoding by the \gls{psl} at the receiving side results in no activated synapse (i.e.,  \textit{continue} in line 3 of \algoref{alg:psl_simp} always triggered).
The system functions correctly without an extension. However, transmitted events yielding no activated synapses after decoding increase the energy and processing requirements unnecessarily. Such events are best filtered out by the \gls{scu} in the source already.

This \gls{scu} power-performance optimization is achieved by calculating not only the  XY start-coordinate of the projected kernel (i.e., the top-left anchor point) but also the first X and Y coordinates outside the range. These coordinates are $(x_\text{min},y_\text{min})+(\mathit{KW},\mathit{KH})$. One can then check if the kernel overlaps with at least one neuron of the population through simple Boolean operations. If it does not, no event is generated.
To enable this optimization, {axons} must contain the kernel shape as well as the width and height of the destination population. 

Hit detection in the channel coordinate is not required, as all channels in the XY kernel range  are updated at the destination, irrespective of the channel of the firing neuron. Thus, if cutting in XY is the rare exception, hit-detection support might be skipped to reduce cost. However, hit detection has a reasonable hardware cost. 
Hence, hit-detection support is beneficial in most cases. 
\algoref{alg:scu_cuts} shows the pseudo code for the \gls{scu} with support for \gls{fm} fragmentation and hit detection.
\begin{algorithm}[t]
\caption{$\acrshort{scu}$ for a regular convolution with support for  \gls{fm} fragmentation and empty-event skipping.}
\label{alg:scu_cuts}
\begin{algorithmic}[1]
\footnotesize
\REQUIRE {Firing neuron \& value: $x_\text{src}$, $y_\text{src}$, $c_\text{src}$, $v$}
\REQUIRE {Axon set of population: $\mathcal A$}
\STATE{\color{red} // grayed hit-detection only needed with many XY cuts}
\FOR{  $(X_\text{off}, Y_\text{off}, C_\text{off}, \color{gray!75!black}  W, H, \mathit{KW},  \mathit{KH}, \color{black} \mathit{AD}_\text{c}, \mathit{ID}_\text{p}) \in  \mathcal A$ }
\STATE $x_{\min},y_{\min}, c_\text{src} = (x_\text{src},y_\text{src},c_\text{src}) + (X_\text{off},Y_\text{off},C_\text{off})$
\color{gray!75!black}
\STATE $x_{\max},\, y_{\max} = (x_{\min},y_{\min}) + (\mathit{KW},\, \mathit{KH})$ 
\IF{$(x_{\min}<W)\,\&\,(x_{\max}>0)\,\&\,(y_{\min}<H)\,\&\,(y_{\max}>0)$}
\color{black}
\STATE \textit{gen\_event}($\mathit{AD}_\text{c}, \mathit{ID}_\text{p},  x_{\min}, y_{\min}, c_\text{src}$)
\color{gray!75!black}
\ENDIF
\color{black}
\ENDFOR
\end{algorithmic}
\end{algorithm}

\subsection{Down- and Upsampling Support}
In the following, we extend the \gls{scu}-\gls{psl} pair to support upsampling and strided convolutions.
A  kernel stride of $N$ in a convolution implies that the anchor point of the kernel in the regular/output-centric view (cf.~\figref{fig:eb_conv}.a)
is advanced by $N$ in X between two generated output points. Once the end of a line is reached, the anchor point is advanced by $N$ in Y while X is reset to 0.
Consequently, a kernel stride of $N$ in a convolution results in the same extracted \gls{fm} as applying the convolution kernel regularly (i.e., with a stride of 1), 
 followed by downsampling the resulting \gls{fm} by a factor $N$ in X and Y.\footnote{Downsampling by a factor $N$ means here simply keeping only every $N^\text{th}$ row and column.} This "destination-downsampling view" of a convolution stride is used to derive our technique.



The pseudo-code of the \gls{psl} with support for programmable kernel striding
is shown in \algoref{alg:psl_str}.
\begin{algorithm}[t]
\caption{\acrshort{psl} with support for strided convolutions.}
\label{alg:psl_str}
\begin{algorithmic}[1]
\footnotesize
\REQUIRE Event body:  $x_{\min}, y_{\min}, c_\text{src}, v$
\REQUIRE Neuron population \& descriptor:  $\mathbf{P}, D, W, H$
\REQUIRE Trans. weights \& descriptor:  $\mathbf{W}$, $\mathit{KW}$, $\mathit{KH}$, $\mathit{SL}$
\FOR{$\Delta x \in [0,\, \mathit{KW})$}
\STATE $x= x_{\min} + \Delta x$
\STATE \textbf{continue if} $x \not\in [0,\, W)\, \textbf{ or }\, x \pmod{2^\mathit{SL}} \neq 0$
\FOR{$\Delta y \in [0,\, \mathit{KH})$}
\STATE $y= y_{\min} + \Delta y$
\STATE \textbf{continue if} $y \not\in [0,\, H)\, \textbf{ or }\, y \pmod{2^\mathit{SL}} \neq 0$
\STATE $x,y = (x, y) \gg \mathit{SL} $ {\color{red}~~// XY down-sampling} 
\FOR{$c \in [0,\, D)$}
\STATE \textit{update\_neuron}($\mathbf P_{c,x,y}$, $\mathbf W_{c,\Delta x, \Delta y, c_\text{src}}$,  $v$)
\ENDFOR
\ENDFOR
\ENDFOR
\end{algorithmic}
\end{algorithm}
We add a field $\mathit{SL}$, containing the $\log_2$ of the applied kernel stride, to the kernel descriptor. Through $\mathit{SL}$, the \gls{psl} knows which rows and columns to skip as they are gone after down-sampling such that there is no need to update a neuron state. Thus, our stride implementation based on destination downsampling results in no compute overhead as accumulation for neurons removed by downsampling is skipped.

The \gls{scu} still behaves as it would for stride-1 convolutions. Hence, the widths $W$ and heights $H$ of the destination fragments in the axons
are the ones that would be obtained for a stride of 1. Thus, the physical width and height of the true/final downsampled \gls{fm} shifted left by $\mathit{SL}$ ($W_t \ll \mathit{SL}$ and $H_t \ll \mathit{SL}$) are stored in the axon as $W$ and $H$. This inverses the downsampling effects. Also, the offset values are adjusted to account for the destination-\gls{fm} downsampling:
\begin{align}
    X_\text{off}  &= X_\text{$0$,src}- \mathit{KW} +  \mathit{XP} + 1 - (X_\text{$0$,dst} \ll \mathit{SL}) \nonumber \\
    Y_\text{off} &= Y_\text{$0$,src}- \mathit{KH} +  \mathit{YP} + 1 - (Y_\text{$0$,dst} \ll \mathit{SL}) \text{.}
\end{align}
All this only affects constant parameters stored in the {axons}. Thus, the \gls{scu} hardware remains unchanged.

Also, the \gls{psl} initially behaves as for a regular convolution. Thus, also here the true population width and height shifted left by $\mathit{SL}$ are stored in the descriptor as $W$ and $H$, respectively. 
The {effective downsampling} after convolution due to the kernel stride is considered by extending the \textit{continue} conditions. In detail, if the \textit{for-loop} iteration over the transposed kernel is at a row or column removed by the downsampling, synapse generation is skipped.  Rows and columns that are not skipped are only the ones with  $X$ or $Y$ coordinates which are 0 modulo $2^\mathit{SL}$.
For non-skipped kernel parts, the downsampling of the XY coordinate is applied for the selection of the neuron to be updated. This down-sampling is realized in hardware at a low cost through a right shift by $\mathit{SL}$. 

Adding stride support does not add noticeably to the complexity of the \gls{psl}\@. First, checking that a binary signal is not 0 in modulo $2^\mathit{SL}$ just requires a logical \textit{OR} operation of the last $\mathit{SL}$ bits. Moreover, all common \glspl{cnn} only use strides of 1 or 2. Thus, we propose to keep $\mathit{SL}$ a 1-bit value such that a stride of 1 (regular convolution) or 2 can be applied in a single step. Larger strides can still be implemented  by inserting dummy/identity layers with another stride of two until the needed downsampling rate is reached. This follows the architectural principle of optimizing for the common case, with non-optimized support for the uncommon case.
With a 1-bit $\mathit{SL}$ field, checking that a signal is not 0 modulo $2^\mathit{SL}$, requires only a single \textit{AND}-gate 
with the $\mathit{SL}$ bit and the signal's \gls{lsb} as the two binary inputs. 
Moreover, with a 1-bit $\mathit{SL}$ value, a shift left by $\mathit{SL}$ has a negligible implementation cost. It can be synthesized in a single 2-to-1 multiplexer with $\mathit{SL}$ as the select signal. Thus, the hardware cost for adding stride support is low.

Next, we discuss the \gls{scu} changes to support upsampling the source before a convolution operation is applied. 
This is required for example for fractionally-strided convolutions~\cite{ronneberger2015unet}. Upsampling  requires filling zeros between the upsampled source values.
Source upsampling is realized in the \gls{scu} by shifting the XY coordinates of the firing neuron left by the $\log_2$ of the upsampling factor $\mathit{US}$ before adding the offset values. Also, the X and Y offsets calculated at compile-time must consider the upsampling of the source:

\begin{align}
     X_\text{off}  &= (X_\text{$0$,src} \ll \mathit{US}) - \mathit{KW} + \mathit{XP} + 1 - (X_\text{$0$,dst} \ll \mathit{SL}) \nonumber \\
    Y_\text{off} &= (Y_\text{$0$,src}  \ll \mathit{US})- \mathit{KH} + \mathit{YP} + 1 - (Y_\text{$0$,dst} \ll \mathit{SL}) 
    \text{.}
\end{align}
The resulting final pseudo code of the  \gls{scu} with upsampling support is shown in \algoref{alg:scu_cuts_upsamp}.
\begin{algorithm}[t]
\caption{$\acrshort{scu}$ with upsampling support.}
\label{alg:scu_cuts_upsamp}
\begin{algorithmic}[1]
\footnotesize
\REQUIRE {Firing neuron \& value: $x_\text{src}$, $y_\text{src}$, $c_\text{src}$, $v$}
\REQUIRE {Axon set of population: $\mathcal A$}
\STATE{\color{red} // grayed hit-detection only needed with many XY cuts}
\FOR{  $(X_\text{off}, Y_\text{off}, C_\text{off},  \color{gray!75!black} W, H, \mathit{KW},  \mathit{KH}, \color{black} \mathit{US}, \mathit{AD}_\text{c}, \mathit{ID}_\text{p})  \in  \mathcal A$ }
\STATE $x_\text{src},y_\text{src} = (x_{\text{src}}, y_{\text{src}}) \ll \mathit{US} $ {\color{red}~~// XY up-sampling}
\STATE $x_{\min},y_{\min}, c_\text{src} = (x_\text{src},y_\text{src},c_\text{src}) + (X_\text{off},Y_\text{off},C_\text{off})$
\color{gray!75!black}
\STATE $x_{\max},y_{\max} =
(x_{\min},y_{\min}) + (\mathit{KW} + \mathit{KH})$ 
\IF{$(x_{\min}<W)\,\&\,(x_{\max}>0)\,\&\,(y_{\min}<H)\,\&\,(y_{\max}>0)$}
\color{black}
\STATE \textit{gen\_event}($\mathit{AD}_\text{c}, \mathit{ID}_\text{p},  x_{\min}, y_{\min}, c_\text{src}$)
\color{gray!75!black}
\ENDIF
\color{black}
\ENDFOR
\end{algorithmic}

\end{algorithm}

\section{Discussions \& Experimental Results}\label{sec:exp_res}
In this section, the proposed technique is evaluated through experimental results. Furthermore, we discuss the universal usability of the technique.

As stated before, the proposed synapse compression technique is \textit{lossless}. Hence, it has no impact on the accuracy or prediction quality of the \glspl{cnn} to be executed. Thus, this section does not discuss network accuracies as the accuracies reported in the original papers of the investigated \glspl{cnn} can be matched. 

The derivation of our proposed technique only considered some of the layer types/operations found in modern \glspl{cnn}\@. Thus, this section first discusses how the technique is used to implement the various additional layer types\@.
Afterwards, we discuss silicon-implementation results. Finally, we determine the compression gains for state-of-the-art \glspl{cnn} through experimental results.

\subsection{Implementing Various Layer-Types}
Today's \glspl{cnn} contain various layer types beyond the ones that were considered for the derivation of the proposed technique in \secref{sec:tech}. Still, our technique based on a \gls{scu}-\gls{psl} pair is able to support them, as shown in the following for the most relevant examples.

\paragraph*{\textbf{Deconvolutions (Transposed Convolutions)}} 
Deconvolutions are used to inverse convolution operations. They are typically implemented as a transposed convolution operation shown in \figref{fig:eb_conv}.b with the addition that the source padding and destination \gls{fm} size is adjusted such that the transposed kernel always fully overlaps with the destination \gls{fm}\@.  Thus, transposed convolutions are  naturally supported by the proposed technique. 

\paragraph*{\textbf{Concatenation \& Split Layers}}
Some neural networks have multiple branches that are at some point concatenated into one. This concatenation and splitting can be implemented through the \gls{scu} due to the support of \gls{fm} fragmentation, described in \subsecref{sec:technique_frag}.

\paragraph*{\textbf{Dilated Convolutions}}
Dilated convolutions are used to increase the receptive field of the kernel without increasing the number of trainable parameters. The idea is to add holes/zeros between the trainable weights of the kernel. The number of skipped locations in X or Y between two trainable weights is called the \textit{dilation rate}.
Hardware support for efficient dilated convolutions can be easily added. However, dilated convolutions are not found in most modern \glspl{cnn}\@. Hence---based on the architectural principle to only optimize for the common case---we propose to implement dilated convolutions, simply as regular convolutions.  If $\mathit{DR}$ is the dilation rate, the used regular convolution kernel has an XY shape of
  $(\mathit{DR}\cdot\mathit{KW}-\mathit{DR}+1) \times (\mathit{DR}\cdot\mathit{KH}-\mathit{DR}+1)$. In this kernel, zeros are inserted at the intermediate positions between the learned weights. 
 Ideally, this is paired with an efficient zero-weight skipping technique as it is done in our silicon implementation presented in the next subsection. Note that zero skipping furthermore enables an additional drastic weight compression through pruning. Thus, zero-skipping support is typically anyway desired~\cite{han2016eie}.

\paragraph*{\textbf{Depthwise \& Grouped Convolutions}}
A depthwise convolution is a lightweight filtering, applying a single convolution per input channel (i.e., no channel mixing; $i^\text{th}$ output channel only depends on $i^\text{th}$ input channel instead of all).
 In a grouped convolution, each output channel depends on a small set of $D_\text{group}$ input channels.
Depthwise and grouped convolutions can be implemented with the \gls{scu}-\gls{psl} pair by splitting the original source and destination \gls{fm} into many \glspl{fm} of depth 1 (depthwise) or $D_\text{group}$ (grouped), with regular convolution operations between the source-destination pairs covering the same channel(s). 




\paragraph*{\textbf{Average \&  Max.~Pooling}}
Pooling operations are implemented as strided depthwise convolutions. For example, an average pooling over non-overlapping $2\times 2$ windows is simply  
a stride-2, depthwise, $2\times2$ convolution. Thereby, all four weights are $\nicefrac{1}{4}$. Max.~pooling has the same connectivity. The only difference is that the weights are 1 rather than $\nicefrac{1}{\text{PoolingWindowSize}}$. 
Thus, only the neuron-update routine is changed for max.~pooling, not the proposed synapse compression technique.

\paragraph*{\textbf{Dense Layers}}
A fully-connected layer from $N$ to $N^+$ neurons is the same as a $1 \times 1$ convolution between two \glspl{fm} of shape $N\times 1 \times 1$ and $N^+\times 1 \times 1$. This is naturally supported by the proposed techniques. In this case, each neuron forms an individual {feature}. Thus, all neurons in the destination population depend on all neurons in the source population.

\paragraph*{\textbf{Flattening \& Global Pooling}}
Flattening an \gls{fm} of shape $D\times W \times H$ followed by a dense layer with $N$ neurons  (i.e., destination \gls{fm} shape is $N\times 1 \times 1$) is implemented as a single regular convolution with a kernel XY shape equal to $(W,H)$ with $N$ output channels. Thereby, the two layers are merged into a single operation which saves compute and events.
Global pooling is realized in the same fashion, only that a depthwise connectivity must be implemented.

\paragraph*{\textbf{Nearest-Neighbor \& Bilinear Interpolation}}
An upsampling together with an interpolation  requires combining the upsampling feature with an untrainable depthwise convolution representing the interpolation between the upsampled points. 

\paragraph*{\textbf{Add and Multiply Layers}}
Add layers and pointwise multiply layers  require two individual source \glspl{fm} of the same shape pointing to the same destination \gls{fm}\@. 
Between the two individual source \glspl{fm} and the destination \gls{fm} the same pointwise synaptic connectivity is required. This is implemented as the connectivity of a  depthwise $1\times 1$ convolution with a weight of 1 shared by both source \glspl{fm}\@. 

\subsection{Silicon Implementation}
A variant of the \gls{scu}-\gls{psl} pair supporting all layer types
is implemented 
in our next-generation \textit{GrAI} architecture, taped-out in the \gls{tsmc}12 \gls{finfet} technology. The heart of the  \unit[60]{mm$^2$} \gls{soc} (shown in \figref{fig:die_shot} on page \pageref{fig:die_shot}) is a self-contained, event-based accelerator, supporting standard \glspl{dnn}, \glspl{lif}, and \glspl{sd} with over 18 million neurons. 

The individual cores are arranged as an $XY$
mesh connected through an \gls{noc}\@. 
By using an 8-bit relative core address (4-bit X, 4-bit Y), a neuron can have a synaptic connection with any 
neuron in the same core or any of the nearest 255 cores, while enabling the architecture to theoretically scale to an arbitrary number of cores. Our pre-product chip for embedded applications has 144 cores per die, but chip tiling---enabled by the relative addressing scheme---still allows for meshes with larger core counts.

Due to the self-contained nature, all neurons, weights, and other parameters mapped to a core must fit into the on-chip \glspl{sram}\@. 
Each core contains \unit[256]{kB} of unified local memory (with 64-bit words and 15-bit addresses) that can be allocated freely to weights, neuron states, and connectivity.

To utilize the memory optimally, the architecture uses 8 bits to describe the width and height of a neuron population and 10 bits to describe the depth. This is enough for modern \glspl{cnn}\@. Larger populations can still be realized through \gls{fm} cutting. The addressing allows a single population to fit over 1 million neurons. Still a neuron-population descriptor containing width, height, depth, neuron type, activation function, {axon} count, and start address of the population in the memory  fits in one 64-bit word. Note that in our silicon implementation {axons} and states are stored in a continuous block. Thus, we can store only the axon count instead of a 15-bit pointer in the population descriptor, which 
saves bits.

The kernel width, $\mathit{KW}$, and height, $\mathit{KH}$, in the {axons} and kernel descriptors are 4-bit numbers. The selection of the bit widths used for kernel width and height are important design choices. Together with the width of an XY coordinate (here 8 bit), they determine the width of most data-path components in \gls{psl} and \gls{scu}\@. Moreover, they determine the maximum range of the \textit{for loops} implemented through the control path. 

We found that rarely a \gls{cnn} has a kernel size larger than \textit{16}. Hence, the used 4 bits do not only yield a low hardware complexity but also an efficient support of most \glspl{cnn}\@. Note that, due to the flexibility provided by the offset fields $X_\text{off}$ and $Y_\text{off}$, any larger kernel size can still be implemented through multiple {axons} and weight matrices. For example, a $32\times 16$ convolution is realized as a $16\times 16$ convolution paired with another $16\times 16$ convolution between the same \glspl{fm} for which the $X_\text{offset}$ is increased by 16. 
The upsample and stride fields are 3 bits and 1 bit wide, respectively---enough for all common \glspl{cnn}\@. Nevertheless, larger up- and downsample factors can be realized by adding dummy layers with an identity weight matrix. 
 
 With 8-bit XY coordinates and 4-bit kernel dimensions,  $X_\text{off}$ and $Y_\text{off}$ are 9-bit signed values. We constrain \gls{fm} fragments created by the mapper to have a width and height of at least 8 neurons (except for the last remaining cuts towards the left and bottom). This allows reducing the bit width for $W$ and $H$ in the \textit{axons} and the hardware complexity of the hit detection.  In practice, this implies no limitation. The reason is that the fragmentation by channels is not limited, which is strongly preferred anyway.
 
 Given all that, also an {axon} and a kernel descriptor fit comfortably into a single 64-bit word. We made the design choice to have a kernel descriptor per channel of the source \gls{fm} instead of only one per \gls{fm}. Thus, $c_\text{src}$ of the event is used besides $\mathit{ID}_\text{p}$ to select the kernel descriptor.
 A kernel descriptor contains the \gls{3d} kernel shape $(\mathit{KD},\mathit{KW},\mathit{KH})$, and a pointer to the start address of the  resulting \gls{3d} sub-weight-matrix, $\mathbf{W}_{\text{sub},c_\text{src}} = \mathbf{W}_{i,j,k,l}|_{l=c_\text{src}}$, in the memory. Moreover, the kernel descriptor contains information required for zero-weight skipping and weight quantization, which is not relevant to the synapse-compression technique.
 
 Multiple kernel descriptors per layer enable weight reuse among different source channels as well as different quantization and pruning schemes for each sub-weight matrix. Also, it overcomes the need to calculate the start point in the weight matrix the descriptor points at. All weights in the sub-matrix are traversed on an event processing. The drawback is a slightly reduced compression as we need at least $C_\text{src}$ rather than one kernel descriptor per layer.
 
Each neuron can have a persistent state for \gls{sd} or \gls{lif} inference, or can dynamically allocate a temporary accumulator state for regular \gls{dnn} implementations. Dynamic state allocation allows mapping \glspl{cnn} with larger neuron counts, as state entries can be shared among neurons whose accumulation phases are mutually exclusive. A state is a 16-bit float number for maximum precision, while a weight can be quantized to 8 bits or even less with no noticeable performance loss using adaptive-float quantization~\cite{tambe2019adaptivfloat}. 
 
 Other than that, the architecture supports multi-threading, \gls{simd} execution, as well as a wide range of neuron types 
 and activation functions. Also,  \gls{psl} and \gls{scu} still support absolute/non-compressed synapses for smaller network layers with irregular connectivity.
 
 \subsubsection{Impact on Power, Performance, and Area}
 In the following, we summarize the results of the physical implementation in the 12-nm technology. 
 The \gls{psl} and \gls{scu} take in total less than \unit[2]{\%} of the area of a processor node. Only a marginal fraction ($\ll 1\%)$ of the power consumption is due to the \gls{psl} or the \gls{scu}, which are also not found to be part of the critical timing path.
Note that the implemented \gls{psl} and \gls{scu} include many additional features beyond the scope of this paper. Examples are support for multi-threading, weight quantization plus zero skipping, and non-compressed synaptic connectivity. Thus, the hardware costs of the proposed technique are negligible. 
 
 In line with previous work~\cite{han2015deep}---and despite applying the synapse compression---area, power, and timing of the chip are dominated by the on-chip memories used to store the parameters, neuron states, and connectivity. The memories take over 70{\%} of the core area as shown in \figref{fig:die_shot} on page \pageref{fig:die_shot}. Thus, the power and area savings of the proposed technique are determined by the memory savings, investigated in the next subsection.

 \subsection{Compression Results}
Through a Python tool, the memory savings of the proposed synapse-compression technique are quantified.
 The tool calculates the synaptic memory  requirements of a given \gls{cnn} for the proposed method, a simple \gls{lut} approach, and the hierarchical-\gls{lut} approach~\cite{moradi2017scalable,davies2018loihi}.
 For the proposed technique---other than for the reference techniques---the memory requirements of a layer increase with the number of fragments the layer is split in. Thus, the tool calculates the memory requirements for the proposed scheme considering \gls{fm} cuts that ensure that the total memory footprint of each resulting fragment is below \unit[256]{kB}, the single core limit of our chip. By considering the real (non-beneficial) required \gls{fm} cuts for the proposed technique, we report realistic rather than optimistic gains for the proposed technique.
 
 To enable a detailed analysis, the memory requirements are categorized into \textit{neurons, connectivity}, and \textit{parameters}. 
 The reported values for \textit{connectivity} are the bits required to store the connectivity between the neurons, but they do not include the weights of the synaptic connections. Thus, for the proposed technique, it includes {axons}, kernel descriptors, and population descriptors.
 The category \textit{parameters} includes the weights. The last section, \textit{neurons}, reports the memory allocated for neuron states. Throughout this section, we assume that a persistent 16-bit state is allocated for each neuron in a population. 
 Reporting the memory usage in these categories allows to access the gains of pattern/weight sharing and {axon}-based synapse computation in isolation. The former technique reduces the \textit{parameter} requirements, while the latter reduces the \textit{connectivity} requirements.
 
In the experiments, the tag width in the reference hierarchical-\gls{lut} technique is 15 {bits}. This is the minimum tag width required by the scheme to be able to map all \glspl{cnn} analyzed in this section, as the analyzed networks have neurons with a fan-in (i.e., number of incoming synapses) of up to $7^2\cdot 512$. 
Thereby, we present best-case values for the hierarchical-\gls{lut} reference technique, which in this form is only able to support \glspl{cnn} with synaptic fan-ins of at most \unit[32]{k}.

 The unified memory can theoretically fit $2^{17}$ 16-bit neuron states in $\unit[256]{kB}$ of unified memory. Still, for the naive \gls{lut} technique, we consider only 15-bit neuron addresses, as none of our experiments resulted in a core memory that is occupied more than \unit[25]{\%} by neurons for this technique. Hereby, we ensure a fair comparison also for the second reference technique. \tableref{tab:bit_width} summarises again the bit widths of all data types for the three techniques.

 \begin{table}[t]
\centering
\scriptsize
\setlength{\tabcolsep}{2.3pt} 
\caption{Bit width of instances of the data types in the various synapse compression techniques used for the experimental results.}\label{tab:bit_width}
\begin{tabular}{cccc}
\toprule
 \multicolumn{1}{c}{\textbf{Technique}} &  \multicolumn{1}{c}{\textbf{Neurons}}  & \multicolumn{1}{c}{\textbf{Parameters}} & \multicolumn{1}{c}{\textbf{Connectivity}} \\ \toprule
                           \multirow{3}{*}{This work}          &  \multirow{3}{*}{\unit[16]{b}}  &  \multirow{3}{*}{\unit[8]{b}} &  \textbf{Axons:} \unit[64]{b}              \\
                                                               &                                 &                               &  \textbf{Kernel Descriptor:} \unit[64]{b}      \\
                                                                  &                                 &                               &  \textbf{Population Descriptor:} \unit[64]{b}      \\
 \midrule                                                              
                            \multirow{1}{*}{\gls{lut}}          &  \multirow{2}{*}{\unit[16]{b}}  &  \multirow{2}{*}{\unit[8]{b}} &  \textbf{\gls{lut} Entry:} \unit[23]{b} \\ 
                            \multirow{1}{*}{\cite{bojarski2016end}} & & & \textit{(\unit[8]{b} Core Address \ + \unit[15]{b} Neuron ID)}             \\
\midrule

                                                              &  \multirow{4}{*}{\unit[16]{b}}  &  \multirow{4}{*}{\unit[8]{b}} &  \textbf{Source~\gls{lut} Entry:} \unit[23]{b}             \\
                           \multirow{1}{*}{Hier.~\gls{lut}}   &                                 &                               & \textit{(\unit[8]{b} Core Address\ + \unit[15]{b} Tag)}  \\  
                           \multirow{1}{*}{\cite{moradi2017scalable,davies2018loihi}}           &                                 &                               &  \textbf{Destination~\gls{lut} Entry:} \unit[15]{b} \\ & & & \textit{(Neuron ID)}      \\

 \bottomrule 
\end{tabular}
\end{table}

By considering persistent 16-bit states for each neuron but 8-bit weights/parameters, we present pessimistic
compression gains for the proposed technique. Our technique heavily compresses the \textit{parameters} and the \textit{connectivity} compared to the reference techniques but not the \textit{neurons}.
Generally, both, the effectively required
bits per neuron and per parameter can be reduced further by compression techniques orthogonal to the one proposed in this work (e.g., non-persistent states, weight pruning, entropy coding, channel pruning)~\cite{han2015deep,he2017channel}. However, reducing the memory footprints of neurons and parameters would automatically increase the compression gains of our proposed technique. The heavily compressed connectivity in our scheme would have a higher impact in this case.
 
Through this pessimistic experimental setup, we ensure that the reported gains of our technique are realistically achieved in practically all applications of event-based architectures.  On the downside, the setup results in pessimistic total memory requirements, why the reported values allow no final conclusion on the mappability of the analyzed \glspl{cnn} on our chip. 
Thus, the reported memory requirements are above what is realistically achieved in our chip, which also supports many parameter and neuron compression schemes orthogonal to the proposed technique. 

\subsubsection{Small CNNs}
We start by analyzing a relatively small \gls{cnn} for today's standards, \textit{PilotNet}~\cite{bojarski2016end}. It is a 9-layer \gls{cnn} used for autonomous steering.
 This example is discussed in-depth first, as it allows us to understand the advantages as well as challenges revealed by the proposed technique. Also, \textit{PilotNet} is the network that was demonstrated to fit in \textit{Intel}'s newest event-based architecture, \textit{Loihi 2}, making it a good reference benchmark. In the following subsection, we will validate the findings for more modern (and much larger) networks.
 
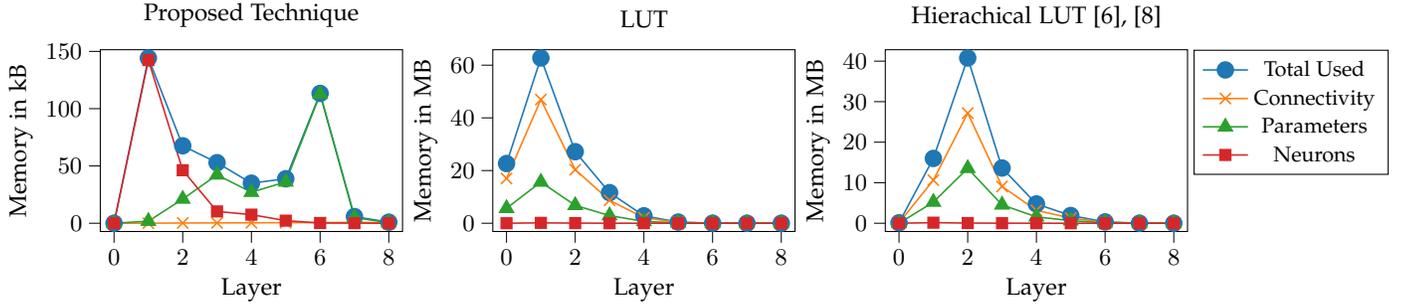
\begin{figure*}
      \centering  
      \input{./tikz/pilotnet_plot}
      \vspace*{-0.2cm}
	    \caption{Memory requirements for \textit{PilotNet}~\cite{bojarski2016end} with the proposed and reference techniques. Note the different Y scale (\unit{kB} vs.\ \unit{MB}).}\label{fig:plot_pilotnet}  
\end{figure*}

We calculate the memory requirements of each layer for the proposed and the two reference techniques. The results are plotted in \figref{fig:plot_pilotnet}.
They show that the memory requirements of the two other published techniques are dominated by \textit{connectivity}. For the two reference techniques \gls{lut} and hierarchical \gls{lut} \unit[74.1]{\%} and \unit[65.2]{\%} of the memory requirements are due to \textit{connectivity}, respectively. The \textit{parameters} are the second dominant factor for the memory requirements of the reference techniques, contributing about \unit[25]{\%} to \unit[35]{\%} to the total memory usage. 
For non-quantized 16-bit weights, the \textit{parameter} requirements would compete with the \textit{connectivity} requirements.

Despite allocating a 16-bit state per neuron, the \textit{neurons} only account for \unit[0.2]{\%} to \unit[0.3]{\%}  of the memory requirements for the reference techniques. This proves that synapse compression---as done by the proposed technique---is the most effective way to reduce the memory requirements for event-based architectures at no accuracy loss.

The proposed synapse-computation methodology drastically reduces the \textit{connectivity} and \textit{parameter} requirements by a factor of \unit[15.6\,k]{$\times$} and \unit[107]{$\times$}, compared to the best previously published method. \textit{Intel} reports for the  upcoming \textit{Loihi 2} architecture a \gls{cnn} synapse compression of up to \unit[17]{$\times$}.\footnote{As of writing this paper, only maximum compression results are available for \textit{Loihi 2}~\cite{loihi2:online}. The details of the used technique are not fully disclosed. Thus, more comparisons with \textit{Loihi 2} cannot be drawn.} The synapses, i.e., the combination of \textit{parameters} and \textit{connectivity}, are compressed by the proposed technique by a factor of \unit[305]{$\times$}. This means an additional compression by a factor of \unit[18]{$\times$} on top of the \unit[17]{$\times$} published by \textit{Intel}~\cite{loihi2:online}.

 \textit{Connectivity}, from being the biggest contributor to the memory requirements, effectively becomes negligible with the proposed technique, contributing only \unit[0.7]{\%} to the memory usage. \textit{Parameters}, on the other hand, are still dominant (\unit[53.7]{\%}), despite the  weight-sharing scheme. \textit{Neurons} start to become an important factor  as well (\unit[45.6]{\%}). This is only due to the persistent neuron states. For stateless neurons, the memory requirements are weight dominated, demonstrating a high memory efficiency of our architecture also for
 stateless neurons and low-bit-width weight quantization.
 
\figref{fig:plot_pilotnet} shows that  the memory requirements of the first layers are dominated by \textit{neurons} with the proposed technique (for the considered stateful neurons). Later layers are \textit{parameter} dominated. This strongly justifies the need for a unified memory where the section sizes for \textit{parameters} and \textit{neurons} can be freely assigned, as done in our taped-out chip. 
The growing ratio of \textit{neuron} over \textit{parameter} requirements is explained with the structure of a typical \gls{cnn}\@. Early layers have a larger XY resolution than later ones. Stride-2 convolutions, non-padded convolutions, and pooling layers successively reduce the width, $W$, and height, $H$, of the \glspl{fm}\@. On the other hand, the channel count, $D$ increases with the depth of the layer in most \glspl{cnn}\@. Since the neuron and weight counts of a regular convolution layer are $D$$\cdot$$W$$\cdot$$H$ and $D_\text{src}$$\cdot$$D$$\cdot$$\mathit{KW}$$\cdot$$\mathit{KH}$, respectively,
weight requirements increase and neuron requirements decrease with the layer depth.

Most layers of the \textit{PilotNet}~\gls{cnn} only take a small portion of the memory of a core  with our technique. Consequently, a mapping aiming at a minimized core count, showed that the  proposed synapse-compression technique allows fitting the full \gls{cnn} without any further optimization in as few as 3 out of the 144 cores. 
With the two reference techniques, the same mapping experiment shows a core-count requirement that is at least \unit[101]{$\times$} higher. 

\subsection{State-of-the-art CNNs}
The reference neuromorphic architectures  do not support sophisticated \glspl{cnn}\@. The reason is that such complex networks do not fit in the local memories due to the large synaptic memory requirements. The proposed technique overcomes this limitation through its large compression rates, as shown in the following.

We analyze the core and memory requirements for four of today's most popular \glspl{cnn}: \textit{Google's MobileNet}~\cite{howard2017mobilenets}, \textit{ResNet50}~\cite{he2016deep}, \textit{Darknet53}~\cite{redmon2018yolov3}, and \textit{ResNet101}~\cite{he2016deep}. The first network is designed to run on mobile devices and is thus more lightweight than the other three designed for less constrained devices such as \glspl{gpu}\@. Hence, \textit{MobileNet} has far fewer neurons and synapses than the other three analyzed \glspl{cnn} (but still many more than \textit{PilotNet}). In particular, \textit{ResNet101} is an extremely complex network, while \textit{ResNet50} and \textit{Darknet53} lay somewhere in the middle. 
We investigate these \glspl{cnn} to quantify the gains of the proposed technique for a broad range of \glspl{cnn}\@. Also, we show that the proposed compression technique enables the execution 
of even complex \glspl{cnn} in a single self-contained chip. 
\begin{table}[t]
\centering
\scriptsize
\setlength{\tabcolsep}{2.3pt} 
\caption{Mapping of state-of-the-art \glspl{cnn}  with the proposed synapse-compression scheme versus prior art (in brackets the compression rates of the proposed scheme compared to prior art).}\label{tab:exp_res}
\begin{tabular}{cccccccc}
\toprule
\multicolumn{1}{c}{\textbf{\acrshort{cnn}}} & \multicolumn{1}{c}{\textbf{Syn.~Compr.}} &  \multicolumn{1}{c}{\textbf{Mem.~Total}} &  \multicolumn{1}{c}{\textbf{Neurons}} & \multicolumn{1}{c}{\textbf{Connectivity}} & \multicolumn{1}{c}{\textbf{Parameters}} \\ \toprule
 \multirow{2}{*}{PilotNet}                            & This Work          &  \unit[0.45]{MB}& \unit[0.20]{MB}& \unit[3.16]{kB}  &  \unit[0.24]{MB}      \\
  \multirow{2}{*}{\cite{bojarski2016end}} & \multirow{2}{*}{\gls{lut}}  &  \unit[0.11]{GB} &  \unit[0.20]{MB}& \unit[91.44]{MB} &  \unit[25.63]{MB} \\
   &  &  \textit{(262$\times$) }&  \textit{(1$\times$) }& \textit{(29.6k$\times$) } &  \textit{(107$\times$) } \\
 & \multirow{2}{*}{Hier.~\gls{lut}}  &  \unit[74.30]{MB} &  \unit[0.20]{MB}&  \unit[48.46]{MB} &  \unit[25.63]{MB} \\ 
 &  &  \textit{(166$\times$) }&  \textit{(1$\times$)}& \textit{(15.6k$\times$)} &  \textit{(107$\times$)} \\
 \midrule
\multirow{2}{*}{MobileNet} &  This Work          & \unit[11.23]{MB} & \unit[8.00]{MB} & \unit[0.16]{MB}  & \unit[3.07]{MB}      \\
\multirow{2}{*}{\cite{howard2017mobilenets}}  & 
\multirow{2}{*}{\gls{lut}}  &  
\unit[1.85]{GB}& \unit[8.00]{MB}&  \unit[1.38]{GB} &  \unit[0.46]{GB} \\
& & \textit{(169$\times$)}& \textit{(1$\times$)}&  \textit{(8.9k$\times$) } &  \textit{(154$\times$)} \\
 & \multirow{2}{*}{Hier.~\gls{lut}}   &  \unit[1.34]{GB} &   \unit[8.00]{MB}&  \unit[0.88]{GB}& \unit[0.46]{GB} \\ 
& & \textit{(123$\times$)}& \textit{(1$\times$)}&  \textit{(5.7k$\times$) } &  \textit{(154$\times$)} \\ \midrule
\multirow{2}{*}{ResNet50} & This Work          &   \unit[43.48]{MB} & \unit[17.71]{MB} & \unit[1.31]{MB} & \unit[24.45]{MB}      \\
 \multirow{2}{*}{\cite{he2016deep}} & 
 \multirow{2}{*}{\gls{lut}}  & \unit[14.60]{GB}&  \unit[17.71]{MB}&  \unit[11.03]{GB} & \unit[3.54]{GB} \\
 & & \textit{(344$\times$)}& \textit{(1$\times$)}&  \textit{(8.6k$\times$) } &  \textit{(149$\times$)} \\
 & \multirow{2}{*}{Hier.~\gls{lut}}
 & \unit[10.26]{GB} &   \unit[17.71]{MB}& \unit[6.70]{GB} & \unit[3.54]{GB}  \\ 
  & & \textit{(242$\times$)}& \textit{(1$\times$)}&  \textit{(5.2k$\times$) } &  \textit{(149$\times$)} \\
 \midrule
\multirow{2}{*}{DarkNet53}  & This Work          &  \unit[51.21]{MB}&   \unit[21.19]{MB}& \unit[1.36]{MB} &  \unit[28.66]{MB}      \\
\multirow{2}{*}{\cite{redmon2018yolov3}} & 
\multirow{2}{*}{\gls{lut}}  & \unit[25.13]{GB}&  \unit[21.19]{MB}& \unit[18.63]{GB} &  \unit[6.48]{GB} \\
  & & \textit{(502$\times$)}& \textit{(1$\times$)}&  \textit{(14.0k$\times$) } &  \textit{(231$\times$)} \\ 
 & \multirow{2}{*}{Hier.~\gls{lut}}
 & \unit[18.68]{GB}&  \unit[21.19]{MB}&  \unit[12.18]{GB}  &\unit[6.48]{GB} \\
   & & \textit{(374$\times$)}& \textit{(1$\times$)}&  \textit{(9.2k$\times$) } &  \textit{(231$\times$)} \\ 
 \midrule
\multirow{2}{*}{ResNet101} & This Work          &  \unit[72.23]{MB}& \unit[27.47]{MB}& \unit[2.18]{MB} &  \unit[42.57]{MB}      \\
 \multirow{2}{*}{\cite{he2016deep}} & 
 \multirow{2}{*}{\gls{lut}}  &  \unit[28.02]{GB}& \unit[27.47]{MB}& \unit[20.98]{GB} &  \unit[7.01]{GB} \\
   & & \textit{(397$\times$)}& \textit{(1$\times$)}&  \textit{(9.8k$\times$) } &  \textit{(169$\times$)} \\ 
 & \multirow{2}{*}{Hier.~\gls{lut}} 
 & \unit[20.25]{GB} &  \unit[27.47]{MB}&  \unit[13.21]{GB} & \unit[7.01]{GB} \\ 
    & & \textit{(287$\times$)}& \textit{(1$\times$)}&  \textit{(6.2k$\times$) } &  \textit{(169$\times$)} \\ 
 \bottomrule 
\end{tabular}
\end{table}                
The experimental setup here is the same as for the \textit{PilotNet} analysis in the previous subsection. 

\tableref{tab:exp_res} includes the results for all analyzed networks.
The results show that, with the previously published techniques, mapping \glspl{cnn} more complex than \textit{PilotNet} is out of reach. For each network, the memory requirements for \textit{connectivity} is already far above the reasonable on-chip memory limit. This changes with the proposed technique, which compresses the memory requirements for \textit{connectivity} compared to the best reference technique by up to \unit[15k]{$\times$}.
In fact, for all networks, the \textit{connectivity} requirements become negligibly small. Even for the complex \textit{ResNet101} \gls{cnn}, only \unit[2.18]{MB} are required for \textit{connectivity}. This is far below the on-chip memory limit of modern embedded systems. Thus, in combination with other supported compression techniques such as weight pruning, low bit-width quantization, and stateless execution, even \textit{ResNet101} is mappable for our architecture.

Due to the nature of the experimental setup, \textit{parameters} and \textit{neurons} dominate after applying the proposed synapse compression technique. Still, our weight-sharing approach enables us to bring the \textit{parameter} requirements down by up to \unit[231]{$\times$}. 

Despite the pessimistic experimental setup, our technique achieves overall memory compression rates in the range from \unit[123]{$\times$} to \unit[374]{$\times$} compared to the published state-of-the-art.
Another advantage of the proposed technique is that it has larger gains for complex networks. The memory compression rate for the lightweight \glspl{cnn} \textit{MobileNet} and \textit{Pilotnet}  are the lowest.  For the complex \glspl{cnn} \textit{Darknet53} and \textit{ResNet101}---for which a high compression rate is crucial---the overall compression is about \unit[300]{$\times$}. 

The reason is that complex \gls{cnn} layers have much more channels, resulting in many more synapses per neuron. With an increasing  synapse count per neuron,  the compression rate of the proposed technique grows, as our technique only requires one {axon} to describe the connectivity between two neuron populations, irrespective of the synapse count in between. This increases the mappability of complex \glspl{cnn} on our event-based architecture. 

Finally, we want to draw again an abstract comparison with \textit{Loihi 2}, for which the technology brief  reports a synapse compression of up to 17$\times$ for \glspl{cnn} without disclosing the technique fully~\cite{loihi2:online}.
Compared to the stated 17$\times$ best-case reduction by \textit{Loihi2}, we achieve another factor of at least  \unit[18]{$\times$} (\textit{PilotNet}) and up to \unit[37]{$\times$} (\textit{DarkNet53}) on top. This demonstrates the clear advantage of the proposed technique over not only the published state-of-the-art but also over upcoming commercial architectures.


%
\section{Conclusion}
This work presents a technique to compress the synaptic memory in neuromorphic event-based massive-multicore architectures for efficient \gls{cnn} inference. The technique is based on two lightweight hardware blocks substituting memory costly \acrlongpl{lut}. 
Our proposed technique has been taped-out in a 144-core event-based \gls{cnn} accelerator using a 12-nm technology. The hardware overhead of the proposed technique was found to be negligible. Nevertheless, it demonstrates compression rates for various modern \glspl{cnn} ranging from 123{$\times$} to 374{$\times$}. 

  A systematic combination of the proposed technique with \textit{orthogonal} weight and neuron compression schemes is left for future work. With such techniques, we will be even able to run extremely complex \glspl{cnn} on a small form-factor chip in a temporal-sparse event-based fashion at low power consumption, latency, and cost. 

%% file: tikz/pilotnet_plot.tex
\begin{tikzpicture}
\scriptsize
\definecolor{color0}{rgb}{0.12156862745098,0.466666666666667,0.705882352941177}
\definecolor{color1}{rgb}{1,0.498039215686275,0.0549019607843137}
\definecolor{color2}{rgb}{0.172549019607843,0.627450980392157,0.172549019607843}
\definecolor{color3}{rgb}{0.83921568627451,0.152941176470588,0.156862745098039}

\begin{groupplot}[group style={group size=3 by 1, horizontal sep=1.2cm}]
\nextgroupplot[
height=4.0cm,
tick align=outside,
tick pos=left,
title={Proposed Technique},
width=5.6cm,
x grid style={white!69.0196078431373!black},
xlabel={Layer},
xmin=-0.4, xmax=8.4,
xtick style={color=black},
y grid style={white!69.0196078431373!black},
ylabel={Memory in kB},
ymin=-7.211328125, ymax=151.437890625,
ytick style={color=black}
]
\addplot [semithick, color0, mark=*, mark size=3, mark options={solid}]
table {%
0 0.0078125
1 144.2265625
2 67.59765625
3 52.84375
4 34.953125
5 38.828125
6 113.30859375
7 5.830078125
8 0.919921875
};
\addplot [semithick, color1, mark=x, mark size=3, mark options={solid}]
table {%
0 0.0078125
1 0.0390625
2 0.203125
3 0.296875
4 0.390625
5 0.515625
6 0.515625
7 0.796875
8 0.3984375
};
\addplot [semithick, color2, mark=triangle*, mark size=3, mark options={solid}]
table {%
0 0
1 1.78125
2 21.12890625
3 42.234375
4 27.0625
5 36.0625
6 112.59765625
7 4.931640625
8 0.498046875
};
\addplot [semithick, color3, mark=square*, mark size=2, mark options={solid}]
table {%
0 0
1 142.40625
2 46.265625
3 10.3125
4 7.5
5 2.25
6 0.1953125
7 0.1015625
8 0.0234375
};

\nextgroupplot[
height=4.0cm,
tick align=outside,
tick pos=left,
title={\gls{lut}},
width=5.6cm,
x grid style={white!69.0196078431373!black},
xlabel={Layer},
xmin=-0.4, xmax=8.4,
xtick style={color=black},
y grid style={white!69.0196078431373!black},
ylabel={Memory in MB},
ymin=-3.13599815368652, ymax=65.855961227417,
ytick style={color=black}
]
\addplot [semithick, color0, mark=*, mark size=3, mark options={solid}]
table {%
0 22.6593017578125
1 62.7199630737305
2 27.1539764404297
3 11.6116790771484
4 2.81988525390625
5 0.44171142578125
6 0.0193557739257812
7 0.00205230712890625
8 3.0517578125e-05
};
\addplot [semithick, color1, mark=x, mark size=3, mark options={solid}]
table {%
0 16.9944763183594
1 46.9356536865234
2 20.3315734863281
3 8.701171875
4 2.109375
5 0.32958984375
6 0.0143051147460938
7 0.00143051147460938
8 0
};
\addplot [semithick, color2, mark=triangle*, mark size=3, mark options={solid}]
table {%
0 5.66482543945312
1 15.6452407836914
2 6.77722549438477
3 2.90043640136719
4 0.70318603515625
5 0.10992431640625
6 0.00486373901367188
7 0.000524520874023438
8 9.5367431640625e-06
};
\addplot [semithick, color3, mark=square*, mark size=2, mark options={solid}]
table {%
0 0
1 0.139068603515625
2 0.0451812744140625
3 0.01007080078125
4 0.00732421875
5 0.002197265625
6 0.00019073486328125
7 9.918212890625e-05
8 2.288818359375e-05
};

\nextgroupplot[
height=4.0cm,
tick align=outside,
tick pos=left,
title={Hierachical \gls{lut}~\cite{moradi2017scalable,davies2018loihi}},
width=5.6cm,
x grid style={white!69.0196078431373!black},
xlabel={Layer},
xmin=-0.4, xmax=8.4,
xtick style={color=black},
y grid style={white!69.0196078431373!black},
ylabel={Memory in MB},
ymin=-2.03880653381348, ymax=42.814937210083,
ytick style={color=black},
legend style={at={(1.02,1)},anchor=north west, font=\footnotesize},
]
\addplot [semithick, color0, mark=*, mark size=3, mark options={solid}]
table {%
0 0.113296508789062
1 15.9929122924805
2 40.7761306762695
3 13.6208038330078
4 4.76446533203125
5 1.90399169921875
6 0.330162048339844
7 0.0145950317382812
8 0.00145721435546875
};
\addplot [semithick, color1, mark=x, mark size=3, mark options={solid}]
table {%
0 0.113296508789062
1 10.6387481689453
2 27.1765365600586
3 9.07882690429688
4 3.175048828125
5 1.2689208984375
6 0.220012664794922
7 0.00967979431152344
8 0.00095367431640625
};
\addplot [semithick, color2, mark=triangle*, mark size=3, mark options={solid}]
table {%
0 0
1 5.21509552001953
2 13.5544166564941
3 4.53190612792969
4 1.58209228515625
5 0.63287353515625
6 0.109958648681641
7 0.00481605529785156
8 0.000486373901367188
};
\addplot [semithick, color3, mark=square*, mark size=2, mark options={solid}]
table {%
0 0
1 0.139068603515625
2 0.0451812744140625
3 0.01007080078125
4 0.00732421875
5 0.002197265625
6 0.00019073486328125
7 9.918212890625e-05
8 2.288818359375e-05
};
   \addlegendentry{Total Used}
   \addlegendentry{Connectivity}
   \addlegendentry{Parameters}
   \addlegendentry{Neurons}
\end{groupplot}

\end{tikzpicture}